\documentclass[twocolumn,secnumarabic,amssymb, nobibnotes, aps, pre]{revtex4-2}
\usepackage{amsfonts,amssymb,amsmath} 
\usepackage{color} 
\usepackage{graphicx} 
\usepackage{epstopdf,mathrsfs} 
\usepackage[colorlinks,linktocpage,linkcolor=blue]{hyperref}
\usepackage{multirow}

\begin{document}

\preprint{APS/123-QED}

\title{L\'{e}vy walk revisited: Hermite polynomial expansion approach}


\author{Pengbo Xu$^{1}$}
\author{Weihua Deng$^{1}$}
\author{Trifce Sandev$^{2,3,4}$}

\affiliation{$^1$School of Mathematics and Statistics, Gansu Key Laboratory of Applied Mathematics and Complex Systems, Lanzhou University, Lanzhou 730000, P.R. China}
\affiliation{$^2$Radiation Safety Directorate, Partizanski odredi 143, P.O. Box 22, 1020 Skopje, Macedonia}
\affiliation{$^3$Institute of Physics, Faculty of Natural Sciences and Mathematics, Ss. Cyril and Methodius University, P.O. Box 162, 1001 Skopje, Macedonia}
\affiliation{$^4$Research Center for Computer Science and Information Technologies, Macedonian Academy of Sciences and Arts, Bul. Krste Misirkov 2, 1000 Skopje,
Macedonia}



\begin{abstract}
Integral transform method (Fourier or Laplace transform, etc) is more often effective to do the theoretical analysis for the stochastic processes. However, for the time-space coupled cases, e.g., L\'evy walk or nonlinear cases, integral transform method may fail to be so strong or even do not work again.  Here we provide Hermite polynomial expansion approach, being complementary to integral transform method. Some statistical observables of general L\'evy walks are calculated by the Hermite polynomial expansion approach, and the comparisons are made when both the integral transform method and the newly introduced approach work well.


\end{abstract}

\maketitle

\section{Introduction}

Currently, it is universally acknowledged that anomalous diffusions are ubiquitous in the natural world \cite{gold_2006, metzler_2000}.
In general the diffusion can be classified according to the relation between the mean squared displacement (MSD) and time $t$, that is $\big<x^2 (t)\big>\sim t^\alpha$ with $0<\alpha<1$, $\alpha=1$, and $\alpha>1$ respectively corresponding to subdiffusion, normal diffusion, and superdiffusion \cite{metzler_2000}.
Two popular models of describing anomalous diffusions are random walks \cite{weiss_1994} and Langevin pictures \cite{Coffey2004,DengZhang2019}. The continuous time random walk (CTRW) is composed of two random variables: waiting time and jump length. When the first moment of waiting time and/or second moment of jump length diverge(s), the CTRW usually models anomalous diffusion, and the probability density function (PDF) of positions satisfies the fractional Fokker-Planck equation \cite{metzler_2000,cartea_2007,metzler_2014}. While the CTRW has its equivalent Langevin picture \cite{Fogedby},  the CTRW and Langevin picture also have their own particular advantages or disadvantages in describing anomalous diffusion; for example, it is more convenient to use Langevin picture to model anomalous diffusion under external potential. Sometimes, the Fokker-Planck equation of Langevin picture can be conveniently obtained by means of subordination \cite{magdziarz_2007}.


In the framework of CTRW, if the second moment of the jump length diverges and the average waiting time is finite, the process is L\'{e}vy flight \cite{metzler_2000}. While L\'{e}vy flight can effectively describe many physical phenomena, its divergent second moment implies infinite speed of particles, making the process less physical in some sense. To remedy this weak point of L\'{e}vy flight, the time and space coupled random walk is introduced, named as L\'{e}vy walk \cite{zaburdaev:2015}.
The most popular method to analyze the PDFs of L\'evy walk and CTRW model is integral transform, e.g., Laplace or Fourier transform. However, for L\'evy walk, the drawbacks of this kind of method begin to emerge. For example, the explicit expression of the inverse Fourier transform of the PDF can not be obtained due to the coupled space and time. By using the special properties of Hermite polynomials, we build the Hermite polynomial expansion approach, which can effectively deal with many cases that the integral transform method cannot. So, Hermite polynomial expansion approach is complementary to integral transform method, and in this paper we will use it to discuss more general L\'{e}vy walks.

For the original one dimensional L\'{e}vy walk, in each step, its velocity is taken as a constant $v_0$, the walking time $\tau$ is a random variable obeying a specified distribution, and its motion is symmetric. In fact, the distribution of the direction of motion, as a random variable, plays an important role, especially in higher dimensions \cite{zaburdaev_2016}. The ordinary L\'{e}vy walk generally displays superdiffusion or normal diffusion. One of the conclusions obtained in this paper is that L\'{e}vy walk can also show subdiffusion. A general L\'{e}vy walk introduced in \cite{zaburdaev:2015} is that the velocity itself (not only the direction) can be considered as a random variable \cite{zaburdaev_2008}. Another generalization is the L\'{e}vy walk with multiple internal states \cite{xu_2018}. It seems to be a natural assumption that the larger in distance or longer in time a particle moves in one step, the smaller the velocity is. In \cite{dentz}, the authors take the walking length of each step as $\tau^\alpha$, where $\tau$ is the walking duration of each step. In this paper, we will consider a more general form, in which the velocity is a function of walking distance or walking time in each step, and the equivalence between taking the velocity as a general function of moving distance and as function of moving time will also be built up. Besides, we discuss the L\'{e}vy walk with its velocity depending on the current position; in this case, the integral transform method does not work again and we resort to Hermite polynomial expansion approach.

The rest of the paper is organized as follows. In Sec. \ref{sec 1}, we first establish the Hermite polynomials to approach the L\'evy walk with constant velocity, and verify the new method by comparing the PDF and MSD with the ones obtained by directly performing Fourier and Laplace transforms. Then based on the Hermite polynomial series, the iteration equations for the density of first passage time are given as well. In Sec. \ref{sec2}, we mainly consider the L\'{e}vy walk with its velocity being a function of walking length or walking time of each step. We build the model and analyze it through different ways. According to some specific examples of this kind of L\'evy walk, the interesting phenomena are observed. It turns out that when $v(\rho)=1/\rho$ the L\'{e}vy walk will always show normal diffusion no matter how to choose the walking length distribution. Then in Sec. \ref{sec3}, Hermite polynomial expansion approach is used to deal with the L\'{e}vy walk with its velocity depending on the current position. We conclude the paper with summation in Sec. \ref{sec5}.




\section{Introducing Hermite polynomial expansion approach by analyzing classical L\'evy walk} \label{sec 1}

Taking the classical L\'{e}vy walk as a toy model, we introduce the Hermite polynomial expansion approach and establish its analysis framework. The velocity of the L\'{e}vy walk is taken as a constant denoted by $v_0$. And we consider the one dimensional symmetric case.
Thus for $q(x,t)$--the density of the particle just arriving at position $x$ at time $t$ and having a chance to change the moving direction, there exists the well-known equation
\begin{equation}\label{2.1.1}
\begin{split}
   q(x,t)=&\frac{1}{2}\int_{0}^{t}q(x-v_0 \tau,t-\tau)\phi(\tau)d\tau\\
   &+\frac{1}{2}\int_{0}^{t}q(x+v_0 \tau ,t-\tau)\phi(\tau)d\tau
    +P_0(x)\delta(t),
\end{split}
\end{equation}
where $P_0(x)$ represents the distribution of the initial position and $\phi(\tau)$ still denotes the walking time density. In this section we simply take $P_0(x)=\delta(x)$, where $\delta(x)$ is Dirac delta function. For the density function of finding the particle at position $x$ at time $t$, denoted by $P(x,t)$, 
there is the equation
\begin{equation}\label{2.1.2}
\begin{split}
  P(x,t)=&\frac{1}{2}\int_{0}^{t}q(x-v_0 \tau, t-\tau)\Psi(\tau)d\tau\\
  &+\frac{1}{2}\int_{0}^{t}q(x+v_0 \tau ,t-\tau)\Psi(\tau)d\tau,
  \end{split}
\end{equation}
where $\Psi(\tau)$ represents the survival probability defined as
\begin{equation}
  \Psi(\tau)=\int_{\tau}^{\infty}\phi(\tau')d\tau'.
\end{equation}
Considering that the Hermite polynomials form an orthogonal basis of the Hilbert space with the inner product $\big<f,g\big>=\int_{-\infty}^{\infty} f(x)\bar{g}(x)e^{-x^2}dx$, here we assume that  $q(x,t)$ and $P(x,t)$ can be, respectively, decomposed as
\begin{eqnarray}
  q(x,t) & =\sum_{n=0}^{\infty}H_n(x)T_n(t)\exp(-x^2), \label{2.1.3}\\
  P(x,t) & =\sum_{n=0}^{\infty}H_n(x)\tilde{T}_n(t)\exp(-x^2),  \label{2.1.4}
\end{eqnarray}
where $H_n(x), n=0,1,\cdots$, represent the Hermite polynomials.
Substituting (\ref{2.1.3}) into (\ref{2.1.1}) leads to
\begin{equation}\label{2.1.5}
  \begin{split}
    &\sum_{n=0}^{\infty}H_n(x)T_n(t)\exp(-x^2) \\
       & =\frac{1}{2}\sum_{n=0}^{\infty}\int_{0}^{t}H_n(x-v_0 \tau)\exp[-(x-v_0 \tau)^2]T_n(t-\tau)\phi(\tau)d\tau\\
         &+\frac{1}{2}\sum_{n=0}^{\infty}\int_{0}^{t}H_n(x+v_0 \tau)\exp[-(x+v_0 \tau)^2]T_n(t-\tau)\phi(\tau)d\tau\\
         &+P_0(x)\delta(t).
  \end{split}
\end{equation}
Utilizing the properties of $H_n(x)$ shown in (\ref{App_B_3}) and (\ref{App_B_5}),
multiplying $H_m(x), m=0,1,\cdots,n$ on both sides of (\ref{2.1.5}), and considering
\begin{equation}\label{2.1.7}
\begin{split}
  &\int_{-\infty}^{\infty} H_n(x-v_0 \tau)H_m(x)\exp[-(x-v_0\tau)^2]dx\\ &=\int_{-\infty}^{\infty}H_n(y)\exp(-y^2)H_m(y+v_0 \tau)dy \\
  &=\int_{-\infty}^{\infty}H_n(y)\exp(-y^2)\sum_{k=0}^{m}\frac{m! (2v_0 \tau)^{m-k} H_k(y)}{k!(m-k)!}dy,
\end{split}
\end{equation}
there exists
\begin{widetext}
\begin{equation}\label{2.1.8}
  \sqrt{\pi}2^m m!T_m(t)=\frac{1}{2}\sum_{k=0}^{m}\frac{m!}{k!(m-k)!}\int_{0}^{t}\sqrt{\pi}2^k k!  \big[(2v_0\tau)^{m-k}+(-2v_0\tau)^{m-k}\big]T_k(t-\tau)\phi(\tau)d\tau +H_m(0)\delta(t).
\end{equation}
Taking Laplace transform defined as $\hat{g}(s)=\mathcal{L}_{t\rightarrow s}\{g(t)\}=\int_{0}^{\infty}e^{-s t}g(t)dt$ on both sides of (\ref{2.1.8}), then we obtain the iteration relation of $\{\hat{T}_m(s)\}$ as
\begin{equation} \label{2.1.9}
  \sqrt{\pi}2^m m!\hat{T}_m(s)=\frac{1}{2}\sum_{k=0}^{m}\frac{\sqrt{\pi}2^k m!}{(m-k)!}\big[(2v_0)^{m-k}+(-2v_0)^{m-k}\big]\mathcal{L}_{\tau\rightarrow s}\big\{\tau^{m-k}\phi(\tau)\big\}\hat{T}_k(s)+H_m(0).
\end{equation}
\end{widetext}
Then  $\hat{T}_m(s)$ can be obtained from the iteration relation (\ref{2.1.9}).
Here we show the results of $\hat{T}_0(s)$ and $\hat{T}_2(s)$ as examples, which will be used in the following. Considering $H_0(x)=1$ and some other values of $H_n(x)$ at $x=0$ shown in (\ref{App_B_4}), we obtain
\begin{equation}\label{2.1.10}
  \hat{T}_0(s)=\frac{1}{\sqrt{\pi}(1-\hat{\phi}(s))}.
\end{equation}
Taking $m=2$ in (\ref{2.1.9}) leads to
\begin{equation}\label{2.1.11}
\begin{split}
   2^3\sqrt{\pi} \hat{T}_2(s)  =&\sqrt{\pi}(2v_0)^2 \phi''(s)\hat{T}_0(s)+2^3\sqrt{\pi}\hat{\phi}(s)\hat{T}_2(s) \\
     & +H_2(0).
\end{split}
\end{equation}
Further utilizing (\ref{2.1.10}) results in
\begin{equation}\label{2.1.12}
  \hat{T}_2(s)=\frac{\hat{\phi}(s)+2v_0^2\hat{\phi}''(s)-1}{4\sqrt{\pi}(\hat{\phi}(s)-1)^2}.
\end{equation}
It should be noted that $\hat{T}_m(s)=0$ for odd $m$.

Then we begin to build up the relation between $\hat{T}_m(s)$ and $\hat{\tilde{T}}_m(s)$, which will constitute $P(x,t)$. According to (\ref{2.1.2}) and (\ref{2.1.4}), similarly we have the iteration relation of \{$\hat{\tilde{T}}_m(s)$\} as
\begin{equation}\label{2.1.13}
\begin{split}
   \hat{\tilde{T}}_m(s)=& \frac{1}{2}\sum_{k=0}^{m}\frac{1}{(m-k)!}\big[v_0^{m-k}+
  (-v_0)^{m-k}\big] \\
     & \cdot\mathcal{L}_{\tau\rightarrow s}\big\{\tau^{m-k}\Psi(\tau)\big\}\hat{T}_k(s).
\end{split}
\end{equation}
It can be calculated that
\begin{eqnarray}
  &\hat{\tilde{T}}_0(s)  =\hat{T}_0(s)\hat{\Psi}(s)=\frac{1}{\sqrt{\pi}s}, \label{tilde_T1_OLW}\\
  &\hat{\tilde{T}}_2(s)  =\frac{(\hat{\phi}(s)+2 v_0^2 \hat{\phi}''(s)-1)\hat{\Psi}(s)}{4 \sqrt{\pi}(\hat{\phi}(s)-1)^2}
  +\frac{v_0^2\hat{\Psi}''(s)}{2\sqrt{\pi}(1-\hat{\phi}(s))}.  \label{tilde_T2_OLW}
\end{eqnarray}
According to (\ref{App_B_2}), we have the Fourier transform, which is defined as $\bar{g}(k)=\mathcal{F}\{g(x)\}=\int_{-\infty}^{\infty}e^{-i k x}g(x)dx$,
\begin{equation}\label{2.1.15}
  \mathcal{F}\{H_n(x)\exp(-x^2)\}=\sqrt{\pi}(-ik)^n\exp\bigg(-\frac{k^2}{4}\bigg).
\end{equation}
Thus $\hat{\bar{P}}(k,s)=\sum_{n=0}^{\infty}\sqrt{\pi}(-ik)^n\exp\big(-\frac{k^2}{4}\big)\hat{\tilde{T}}_n(s)$. Since $\hat{\bar{P}}(k=0,s)=\sqrt{\pi}\hat{\tilde{T}}_0(s)=\frac{1}{s}$, we conclude that the PDF is normalized. From (\ref{2.1.13}), it can be noted that $\hat{\tilde{T}}_m(s)=0$ for odd $m$. Therefore
\begin{equation}\label{2.1.16}
\begin{split}
    \bar{P}(k,t)&=\sum_{n=0}^{\infty}\tilde{T}_{2n}(t)(-1)^n\sqrt{\pi}k^{2n}\exp\bigg(-\frac{k^2}{4}\bigg);
\end{split}
\end{equation}
and according to
\begin{equation}\label{2.1.17}
  \big<x^m(t)\big>=(i)^m \frac{d^m}{dk^m}P(k,t)\bigg|_{k=0},
\end{equation}
it can be noted that the odd order moment is $0$, which is reasonable since the process is symmetric and starts at $x=0$. In the following we will verify our results by using them to solve the PDF and MSD of L\'evy walk. According to \cite{zaburdaev:2015}, the Fourier-Laplace transform of the PDF of L\'evy walks has the form
\begin{equation}\label{2.1.14}
\hat{\bar{P}}(k,s)=\frac{\hat{\Psi}(s+i k v_0)+\hat{\Psi}(s-i k v_0)}{2-[\hat{\phi}(s+i k v_0)+\hat{\phi}(s-i k v_0)]}.
\end{equation}
First, it can be simply shown that the first two terms of (\ref{2.1.16}) can also be obtained from (\ref{2.1.14}). In fact, inserting (\ref{tilde_T1_OLW}) and (\ref{tilde_T2_OLW}) into (\ref{2.1.16}) leads to
\begin{equation}\label{LF_pdf_HP}
\begin{split}
   \hat{\bar{P}}(k,s)
   =&\frac{1}{s}\exp\left(-\frac{k^2}{4}\right)
  -k^2\bigg[\frac{(\hat{\phi}(s)+2 v_0^2 \hat{\phi}''(s)-1)\hat{\Psi}(s)}{4
  (\hat{\phi}(s)-1)^2}\\
     &+\frac{v_0^2\hat{\Psi}''(s)}{2(1-\hat{\phi}(s))}\bigg]\exp\left(-\frac{k^2}{4}\right)
      +\sum_{n=2}^{\infty}(-1)^n\sqrt{\pi}k^{2n}\\
      &\cdot\hat{\tilde{T}}_{2n}(s)\exp\bigg(-\frac{k^2}{4}\bigg),
\end{split}
\end{equation}
the first two terms of which can be directly obtained by doing the Taylor expansion of the expression in square bracket at $k=0$ of
\begin{equation}\label{LW_FL_PDF}
\begin{split}
   \hat{\bar{P}}(k,s)= &  \left[\frac{\hat{\Psi}(s+i k v_0)+\hat{\Psi}(s-i k v_0)}{2-[\hat{\phi}(s+i k v_0)+\hat{\phi}(s-i k v_0)]}\exp\left(\frac{k^2}{4}\right)\right] \\
     & \cdot  \exp\left(-\frac{k^2}{4}\right).
\end{split}
\end{equation}

Next we turn to the discussion of MSD.
%
From (\ref{2.1.16}) and (\ref{2.1.17}), there exists
\begin{equation}\label{2.1.18}
  \begin{split}
    & \big<x^2(s)\big> \\
    &=\mathcal{L}_{t\rightarrow s}\{\big<x^2(t)\big>\}\\
       &=\frac{\sqrt{\pi}}{2}\hat{\tilde{T}}_0(s)+2\sqrt{\pi}\hat{\tilde{T}}_2(s)\\
       &=\frac{1}{2s}+\frac{(\hat{\phi}(s)+2 v_0^2 \hat{\phi}''(s)-1)\hat{\Psi}(s)}{2(\hat{\phi}(s)-1)^2}+\frac{v_0^2\hat{\Psi}''(s)}{(1-\hat{\phi}(s))}.
  \end{split}
\end{equation}
Specifically, if we consider the flight time with the power law density, i.e., $\phi(\tau)=\alpha/(\tau_0(1+\tau/\tau_0)^{1+\alpha})$, where $\tau_0>0$ and $\alpha>0$. According to \cite{zaburdaev:2015}, the corresponding asymptotic Laplace transform when $\alpha\neq1,2$ is
\begin{equation}\label{L_powerlaw}
   \hat{\phi}(s)\sim 1-\frac{\tau_0}{\alpha-1}s-\tau_0^\alpha\Gamma(1-\alpha)
   s^\alpha+\frac{\tau_0^2}{(\alpha-2)(\alpha-1)}s^2.
\end{equation}
Substituting (\ref{L_powerlaw})  into (\ref{2.1.18}) recovers the well-know results of long time $t$:
for $0<\alpha<1$, $\big<x^2(t)\big>\sim(1-\alpha)v_0^2t^2$; for $1<\alpha<2$, $\big<x^2(t)\big>\sim\frac{2v_0^2(\alpha-1)}{(3-\alpha)(2-\alpha)}t^{3-\alpha}$; for $\alpha>2$, $\big<x^2(t)\big>\sim\frac{2v_0^2}{\alpha-2}t$.
Another representative flight time distribution is the tempered $\alpha$ stable one, which has the form $c e^{-\lambda x} f_\alpha(x)$ with $0<\alpha<1$ and $\lambda>0$, and $f_\alpha(x)$ represents the one-sided $\alpha$ stable L\'{e}vy distribution. Its Laplace transform is $e^{-[(s+\lambda)^\alpha-\lambda^\alpha]}$\cite{janusz:2010},
the insertion of which into (\ref{2.1.18}) results in: for small $t$ (large $s$),
\begin{equation}\label{2.1.20}
  \big<x^2(s)\big>\sim \frac{2 v_0^2 (1-\alpha \exp(\lambda^\alpha-s^\alpha)s^\alpha)}{s^3}\sim\frac{2v_0^2}{s^3},
\end{equation}
that is $\big<x^2(t)\big>\sim v^2_0 t^2$; while for long time $t$ (small $s$),
\begin{equation}\label{2.1.21}
  \big<x^2(s)\big>\sim \frac{(1 + \alpha (-1 + \lambda^\alpha)) v_0^2}{\lambda s^2},
\end{equation}
that is $\big<x^2(t)\big>\sim \frac{(1 + \alpha (-1 + \lambda^\alpha)) v_0^2 t}{\lambda}$. 
Another recently introduced tempered walking time density is \cite{trifce:2018}
\begin{equation}\label{2.1.22}
  \hat{\phi}(s)=\frac{1}{1+s (s+\lambda)^{\mu-1}[1+(s+\lambda)^{-\delta}]^{\alpha}},
\end{equation}
where $0<\delta<\mu<1$, $0<\alpha<1$, and $\lambda$ is still the tempered parameter. After substituting this walking time distribution into (\ref{2.1.18}), we can still obtain that the MSD transfers from $t^2$ for short time into $t$ for long time.

From the above discussions, one can note that the Hermite polynomial expansion approach solves the issues that the integral transform methods work for. The Hermite polynomial expansion approach can also solve some problems that the integral transform methods can not (or very hard to) deal with, e.g., the density of first passage time.


\subsection{Iteration equations for the density of first passage time}

First passage time is one of the most important statistical quantities \cite{krusemann,deng:2017}. It can be considered as the time that the particle first get out of the domain $\Omega$. From \cite{dybiec}, the following equation connects the density of the first passage time $f(t)$ and the survival probability $S(t):=\int_{\Omega}P(x,t)dx$,
\begin{equation}\label{FPT_S}
  F(t):=\int_{0}^{t}f(u)d u=1-S(t),
\end{equation}
that is $f(t)=-\frac{d}{dt}S(t)$. For L\'evy walk, the density of the first passage time is very hard to calculate, because of the challenge of performing inverse Fourier transform.
However, it can be solved by using Hermite polynomials approach. Here we simply consider the domain $\Omega$ as an interval $[-L,L]$. Then by noticing (\ref{2.1.16}), there exists
\begin{equation}\label{S}
\begin{split}
    S(t)= & \int_{-L}^{L} P(x,t)dx \\
       =&L\sum_{n=0}^{\infty}(-1)^n 2^{1+n} (2n-1)!!\\ &\cdot ~_1F_1\left(\frac{1}{2}+n;\frac{3}{2};-L^2\right)\tilde{T}_{2n}(t),
\end{split}
\end{equation}
where $_1F_1(a;b;z)$ represents the confluent hypergeometric function defined as
\begin{equation}\label{Def_CHF}
  _1F_1(a;b;z)=\sum_{n=0}^{\infty} \frac{a^{(n)}z^n}{b^{(n)}n!}
\end{equation}
with
\begin{equation*}
\begin{split}
   a^{(0)} & =1, \\
   a^{(n)} & =a(a+1)(a+2)\cdots(a+n-1).
\end{split}
\end{equation*}
Thus we conclude that the density of the first passage time $f(t)$ for L\'evy walk satisfies the following equations in the Laplace domain
\begin{widetext}
\begin{equation}\label{L_Den_FPT}
\begin{split}
   \hat{f}(s)&=1-s \hat{S} (s), \\
     \hat{S} (s)  &=  L\sum_{n=0}^{\infty}(-1)^n 2^{1+n} (2n-1)!! _1F_1\left(\frac{1}{2}+n;\frac{3}{2};-L^2\right)\hat{\tilde{T}}_{2n}(s),\\
       \hat{\tilde{T}}_m(s)&=\frac{1}{2}\sum_{k=0}^{m}\frac{1}{(m-k)!}\big[v_0^{m-k}+
  (-v_0)^{m-k}\big]\mathcal{L}_{\tau\rightarrow s}\big\{\tau^{m-k}\Psi(\tau)\big\}\hat{T}_k(s),\\
   \sqrt{\pi}2^m m!\hat{T}_m(s) &= \frac{1}{2}\sum_{k=0}^{m}\frac{\sqrt{\pi}2^k m!}{(m-k)!}\big[(2v_0)^{m-k}+(-2v_0)^{m-k}\big]\mathcal{L}_{\tau\rightarrow s}\big\{\tau^{m-k}\phi(\tau)\big\}\hat{T}_k(s)+H_m(0).
\end{split}
\end{equation}
\end{widetext}
Currently, we only give the equations that the distribution of first passage time satisfies, and in the future, we will further consider how to asymptotically solve these equations.


\section{
L\'{e}vy walk with velocity depending on walking length or walking time of each step
}\label{sec2}

This section focuses on symmetric L\'{e}vy walk with velocity depending on the distance or time of each step, denoted as $\rho$ or $\tau$, respectively. We first build up the models to describe these two kinds of L\'{e}vy walks. Essentially, they are equivalent, which will be shown in the following discussions.

\subsection{L\'{e}vy walk with velocity depending on the distance of each step}\label{1.1}

We consider the symmetric one dimensional L\'{e}vy walk with velocity $v=v(\rho)$ instead of a constant, where $\rho$ represents the walking length of a step with the density $\lambda(\rho)$. Thus the flight time $\tau=\frac{\rho}{v(\rho)}=f(\rho)$ satisfies the density $\phi(\tau)=\lambda(f^{-1}(\tau))\cdot|[f^{-1}(\tau)]'|$ under the assumption that $f(\rho)$ is strictly monotone. Then the density of the particles just reaching position $x$ at time $t$ right after some steps, denoted as $q(x,t)$, satisfies the equation
\begin{equation}\label{Q_Vrho}
  \begin{split}
      & q(x,t)
       \\
        &= \frac{1}{2}\int_{0}^{\infty}d\rho\int_{0}^{t} \big[q(x-\rho,t-\tau)+q(x+\rho,t-\tau)\big]\\
       & ~~~ \cdot \delta(\rho-f^{-1}(\tau))\phi(\tau)d\tau
       +P_0(x)\delta(t),
  \end{split}
\end{equation}
where $P_0(x)\delta(t)$ represents the initial condition. Then we can also find the relation between the PDF $P(x,t)$, representing the probability of finding particles at position $x$ at time $t$, and $q(x,t)$ as
\begin{equation}\label{1.1.5}
  \begin{split}
       & P(x,t)
       \\
       & = \frac{1}{2}\int_{0}^{\infty}d\rho\int_{0}^{t}\big[q(x-\rho,t-\tau)+q(x+\rho,t-\tau)\big]\\
       &~~~ \cdot\delta(\rho-f^{-1}(\tau))\int_{\tau}^{\infty}\phi(\eta) d\eta d \tau.
  \end{split}
\end{equation}
The integral $\int_{\tau}^{\infty}\phi(\eta) d\eta$ in (\ref{1.1.5}) can be equivalently considered as the survival probability in the ordinary L\'{e}vy walk \cite{zaburdaev:2015}.
By the Hermite polynomial expansion approach, inserting (\ref{2.1.3}) into (\ref{Q_Vrho}) leads to
\begin{equation}\label{T_Vrho}
\begin{split}
& \sqrt{\pi}2^m m!\hat{T}_m(s)
  \\
 & = \frac{1}{2}\sum_{k=0}^{m}\frac{\sqrt{\pi}2^k m!}{(m-k)!}\mathcal{L}_{\tau\rightarrow s}\big\{\big[(2f^{-1}(\tau))^{m-k} \\ &~~~+(-2f^{-1}(\tau))^{m-k}\big] \phi(\tau)\big\}\hat{T}_k(s)+H_m(0).
\end{split}
\end{equation}
The $\hat{T}_m(s)$ can be obtained from the above iteration relation, such as,
\begin{equation}\label{T0_T2_vrho}
\begin{split}
   \hat{T}_0(s)&=\frac{1}{\sqrt{\pi}(1-\phi(s))},\\
     \hat{T}_2(s)& =\frac{\mathcal{L}_{\tau\rightarrow s}\{(f^{-1}(\tau))^2\phi(\tau)\}}{2\sqrt{\pi}(1-\phi(s))^2} -\frac{1}{4\sqrt{\pi}(1-\phi(s))}.
\end{split}
\end{equation}
Similarly, the $\{\hat{\tilde{T}}_m(s)\}$ can be got from (\ref{2.1.4}) and (\ref{1.1.5}) as
\begin{equation}\label{TT_vrho}
\begin{split}
 & 2^m  \tilde{T}_m(t)
  \\
  &=   \frac{1}{2} \sum_{k=0}^{m}\frac{2^k}{(m-k)!}\int_{0}^{t} \big[(2f^{-1}(\tau))^{m-k}\\
  & ~~~+(-2f^{-1}(\tau))^{m-k}\big] T_k(t-\tau) \int_{\tau}^{\infty}\phi(\eta)d\eta d\tau.
\end{split}
\end{equation}
Taking Laplace transform of (\ref{TT_vrho}) results in
\begin{equation}\label{LT_TT_vrho}
\begin{split}
  & \hat{\tilde{T}}_m(s)
   \\
   &=\frac{1}{2}\sum_{k=0}^{m}\frac{1}{(m-k)!}\hat{T}_k(s) \mathcal{L}_{\tau\rightarrow s}\bigg\{\big[(-f^{-1}(\tau))^{m-k}\\
   & ~~~ +
  (f^{-1}(\tau))^{m-k}\big]\int_{\tau}^{\infty}\phi(\eta)d\eta\bigg\},
\end{split}
\end{equation}
which leads to
\begin{equation}\label{T0_T2_vrho}
\begin{split}
  \hat{\tilde{T}}_0(s) &=\frac{1}{\sqrt{\pi}s} ,\\
     \hat{\tilde{T}}_2(s)&=\frac{\hat{M}_1}{2\sqrt{\pi}(1-\hat{m}_0)} +\frac{\hat{m}_1}{2\sqrt{\pi}s(1-\hat{m}_0)}-\frac{1}{4\sqrt{\pi}s}
\end{split}
\end{equation}
with
\begin{equation}\label{1.2.19}
  \begin{split}
     \hat{m}_0 &=\mathcal{L}_{\tau\rightarrow s}\{\lambda[f^{-1}(\tau)]|[f^{-1}(\tau)]'|\},\\
      \hat{m}_1 &=\mathcal{L}_{\tau\rightarrow s}\{[f^{-1}(\tau)]^2\lambda[f^{-1}(\tau)]|[f^{-1}(\tau)]'|\},\\
       \hat{M}_1&=\mathcal{L}_{\tau\rightarrow s}\Big\{[f^{-1}(\tau)]^2\int_{\tau}^{\infty}\lambda[f^{-1}(\eta)]|[f^{-1}(\eta)]'|d\eta\Big\}.
  \end{split}
\end{equation}
Then the MSD in Laplace space is
\begin{equation}\label{MSD_vrho}
\begin{split}
  \big<x^2(s)\big>&=\frac{\sqrt{\pi}}{2}\hat{\tilde{T}}_0(s)+2\sqrt{\pi}\hat{\tilde{T}}_2(s)\\
     &  =\frac{\hat{m}_1}{s(1-\hat{m}_0)}+\frac{\hat{M}_1}{1-\hat{m}_0}.
\end{split}
\end{equation}
Next, we derive (\ref{MSD_vrho}) by integral transform method. Taking Fourier transform of (\ref{Q_Vrho}) w.r.t. $x$ leads to
\begin{equation*}
\begin{split}
   \bar{q}(k,t)=&\frac{1}{2}\int_{0}^{t}\big(\exp(i f^{-1}(\tau)k)+\exp(-i f^{-1}(\tau)k)\big)\phi(\tau)\\
   &\cdot\bar{q}(k,t-\tau)d\tau +\bar{P}_0(k)\delta(t)\\
     =&\int_{0}^{t}\cos(f^{-1}(\tau)k)\lambda(f^{-1}(\tau)) |[f^{-1}(\tau)]'|\\
     &\cdot\bar{q}(k,t-\tau)d\tau +\bar{P}_0(k)\delta(t).
\end{split}
\end{equation*}
Further taking Laplace transform w.r.t. $t$ results in
\begin{equation}\label{flt_q_vrho_method2}
\begin{split}
   \hat{\bar{q}}(k,s)=&\mathcal{L}_{\tau\rightarrow s}\{\cos(k f^{-1}(\tau))\lambda(f^{-1}(\tau)) |[f^{-1}(\tau)]'|\}\\
     & \cdot\hat{\bar{q}}(k,s)+\bar{P}_0(k).
\end{split}
\end{equation}
On the other hand, by performing Fourier and Laplace transforms of (\ref{1.1.5}), w.r.t. $x$ and $t$, respectively, we have
\begin{equation}\label{1.1.6}
  \begin{split}
     &\hat{\bar{P}}(k,s)\\
       =& \mathcal{L}_{\tau\rightarrow s}\bigg\{\cos(k f^{-1}(\tau))\int_{\tau}^{\infty}\lambda(f^{-1}(\eta))|[f^{-1}(\eta)]'| d\eta\bigg\}\\
       &\cdot\hat{\bar{q}}(k,s)\\  
       =&\frac{\mathcal{L}_{\tau\rightarrow s}\{\cos(k f^{-1}(\tau))\int_{\tau}^{\infty}\lambda(f^{-1}(\eta))|[f^{-1}(\eta)]'| d\eta\}\bar{P}_0(k)}{1-\mathcal{L}_{\tau\rightarrow s}\{\cos(k f^{-1}(\tau))\lambda(f^{-1}(\tau)) |[f^{-1}(\tau)]'|\}}.
  \end{split}
\end{equation}
\begin{widetext}
From (\ref{1.1.6}), one can easily check the normalization of $P(x,t)$. In fact,
\begin{equation}\label{1.1.10}
  \begin{split}
    \mathcal{L}_{\tau\rightarrow s}\bigg\{\int_{\tau}^{\infty}\lambda(f^{-1}(\eta))|[f^{-1}(\eta)]'| d\eta\bigg\}
 &=\mathcal{L}_{\tau\rightarrow s}\bigg\{1-\int_{0}^{\tau}\lambda(f^{-1}(\eta))|[f^{-1}(\eta)]'| d\eta\bigg\} \\
 &=\frac{1}{s}(1-\mathcal{L}_{\tau\rightarrow s}\{\lambda(f^{-1}(\tau))|[f^{-1}(\tau)]'| \}).
       \end{split}
\end{equation}
Then
\begin{equation}\label{1.1.9}
   \hat{\bar{P}}(k=0,s)=\frac{\mathcal{L}_{\tau\rightarrow s}\{\int_{\tau}^{\infty}\lambda(f^{-1}(\eta))|[f^{-1}(\eta)]'| d\eta\}}{1-\mathcal{L}_{\tau\rightarrow s}\{\lambda(f^{-1}(\tau)) |[f^{-1}(\tau)]'|\}}
 =\frac{1}{s},
\end{equation}
which implies the normalization of $P(x,t)$. Rewriting (\ref{1.1.6}) as
\begin{equation}\label{1.2.1}
  \begin{split}
       \hat{\bar{P}}(k,s) =
      \frac{\sum_{j=0}^{\infty}(-1)^{j}\frac{k^{2j}}{(2j)!}\mathcal{L}_{\tau\rightarrow s}\{(f^{-1}(\tau))^{2j}\int_{\tau}^{\infty}\lambda(f^{-1}(\eta))|[f^{-1}(\eta)]'|d\eta\}\bar{P}_0(k)}
  {1-\sum_{j=0}^{\infty}\frac{k^{2j}}{(2j)!}\mathcal{L}_{\tau\rightarrow s}\{(f^{-1}(\tau))^{2j}\lambda(f^{-1}(\tau))|[f^{-1}(\tau)]'|\}}
  \end{split}
\end{equation}
and taking $P_0(x)=\delta(x)$, the MSD of L\'{e}vy walk can be simply solved, being the same as  (\ref{MSD_vrho}).
\end{widetext}

\subsection{Examples of L\'{e}vy walk with velocity depending on the length of each step}\label{1.2}

In this subsection we will consider some representative velocity functions $v(\rho)$ and calculate the corresponding MSDs. As for $v(\rho)=v_0$, where $v_0$ is a constant, it can be easily verified that (\ref{1.1.6}) reduces to (\ref{2.1.14}). 
In the following, we discuss a little bit general cases, in which $f(\rho)$ is a strictly monotonically increasing function. The request on $f(\rho)$ is reasonable in the sense that the longer distance a particle walks implies the longer time it will take.


\subsubsection{
Being $1/\rho$ for velocity function $v(\rho)$
}\label{sec.3.2.1}

\begin{figure}
  \centering
  \includegraphics[width=8cm]{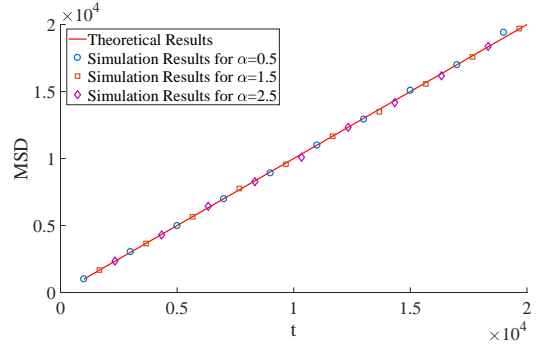}
  \caption{Simulations of MSD for L\'{e}vy walk with velocity $v(\rho)=1/\rho$ and different kinds of walking length distributions $\lambda(\rho)$, sampling over $10^4$ realizations. The dots with different kinds of marks represent the corresponding simulation results of L\'{e}vy walk, whose walking length distribution $\lambda(\rho)=\frac{\alpha}{(1+\rho)^{1+\alpha}}$ with different regions of $\alpha$, while the real line is the theoretical result, $\big<x^2(t)\big>=t$. 
  }\label{fig_1}
\end{figure}

When $v(\rho)$ equals to $1/\rho$, the surprising result is that the MSD always grows linearly with time $t$ no matter what kind of walking length density $\lambda(\rho)$ is.
In this case, $f^{-1}(\tau)=\sqrt{\tau}$.
%
%
Therefore according to (\ref{MSD_vrho}), we have
\begin{equation}\label{add_1.2.2}
\begin{split}
   \hat{m}_1 &= \mathcal{L}_{\tau\rightarrow s}\{\tau \lambda[f^{-1}(\tau)]f^{-1}(\tau)'\}\\
     & =-\frac{d}{ds}\mathcal{L}_{\tau\rightarrow s}\{\lambda[f^{-1}(\tau)]f^{-1}(\tau)'\}\\
     &=-\frac{d}{ds}m_0
\end{split}
\end{equation}
and
\begin{equation}\label{add_1.2.3}
\begin{split}
   \hat{M}_1 &= \mathcal{L}_{\tau\rightarrow s}\bigg\{\tau\bigg(1-\int_{0}^{\tau} \lambda(f^{-1}(\eta))f^{-1}(\eta)'d\eta\bigg)\bigg\} \\
   &=-\frac{d}{ds}\mathcal{L}_{\tau\rightarrow s}\bigg\{1-\int_{0}^{\tau}\lambda(f^{-1}(\eta))f^{-1}(\eta)' d\eta\bigg\}\\
   &=-\frac{d}{ds}\frac{1-m_0}{s}.
\end{split}
\end{equation}
Then
\begin{equation}\label{add_1.2.1}
\begin{split}
     \big<x^2(s)\big> &=\frac{m_1+sM_1}{s(1-m_0)} \\
     &=\frac{-m_0'+\frac{1}{s}(1-m_0+sm_0')}{s(1-m_0)}\\
     & =\frac{1}{s^2},
\end{split}
\end{equation}
which indicates $\big<x^2(t)\big>=t$, meaning normal diffusion.
The result can be verified by the simulations shown in Fig. \ref{fig_1}.

\subsubsection{Being $1/\rho^n$ for velocity function $v(\rho)$}

\begin{figure}
  \centering
  \includegraphics[width=8cm]{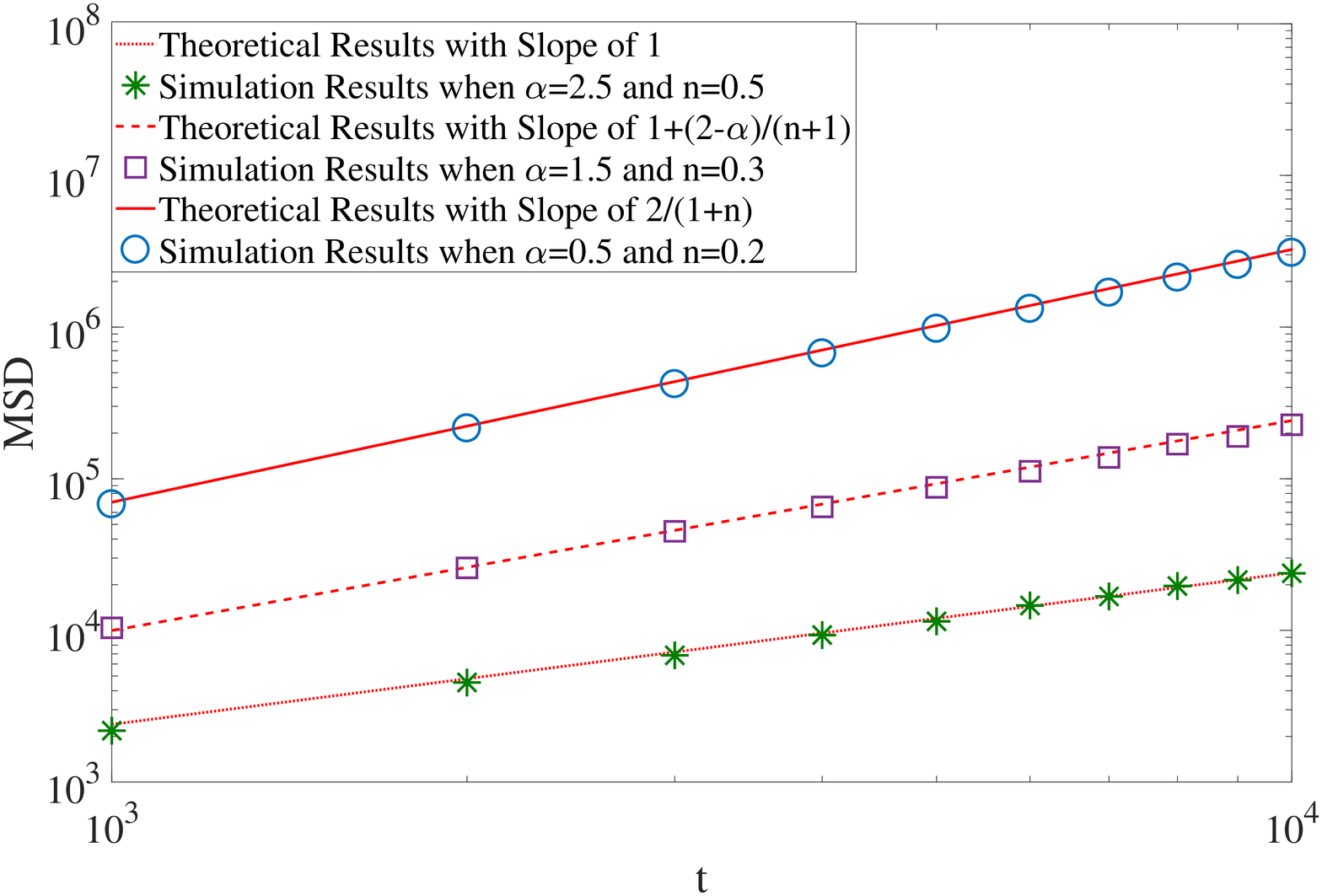}
  \caption{Numerical simulations of MSD of L\'{e}vy walk when $0<n<1$ by sampling over $10^4$ realizations (in log-log scale). 
  }\label{fig_3}
\end{figure}

Consider the general cases of $v(\rho)=\frac{1}{\rho^n}$ with $n>0$. First, let $n\in N$, and then $f^{-1}(\tau)=\tau^{\frac{1}{1+n}}$. The distribution of the walking length is taken as
\begin{equation}\label{1.2.3}
  \lambda(\rho)=\frac{1}{\tau_0}\frac{\alpha}{(1+ \rho/\tau_0)^{1+\alpha}},
\end{equation}
where $\tau_0$ is a constant and in this paper we simply let $\tau_0=1$ for the convenience of calculation. Here we still need to calculate $\hat{m}_0$, $\hat{m}_1$, and $\hat{M}_1$, respectively. According to the definition of $\hat{m}_0$ given in (\ref{1.2.19}), first we need to calculate
\begin{equation}\label{1.2.4}
  \hat{m}_0=\mathcal{L}_{\tau\rightarrow s}\bigg\{\frac{1}{1+n}\tau^{-\frac{n}{1+n}}\frac{\alpha}{(1+\tau^{\frac{1}{1+n}})^{1+\alpha}}\bigg\}.
\end{equation}
According to \cite{prudnikov}, the Laplace transform of (\ref{1.2.4}) can be presented through Meijer G-function defined in (\ref{app_a_1})
\begin{equation}\label{1.2.5}
\begin{split}
   \mathcal{L}_{\tau\rightarrow s}\big\{\tau^{\mu}(\tau^{1/k}+z)^{\nu}\big\}=\bigg(\frac{1}{2\pi}\bigg)^{k-1}\frac{(z/k)^\nu}
  {\Gamma(-\nu)s^{\mu+1}}&\\
      \cdot   G_{k+1,k}^{k,k+1}\bigg(\frac{1}{p}\bigg|
  \begin{matrix}
    \Delta(1,-\mu)&, \Delta(k,\nu+1) \\
    \Delta(k,0)&
  \end{matrix}\bigg)&,
\end{split}
\end{equation}
where ${\rm Re}(\mu)>-1$, $|{\rm arg}(z)|<\pi$, $z\in C$ and $\Delta(k,a)=\frac{a}{k},\frac{a+1}{k},\ldots,\frac{a+k-1}{k}$. Therefore 
\begin{equation}\label{1.2.6}
\begin{split}
    \hat{m}_0=&\frac{\alpha}{1+n}\mathcal{L}_{\tau\rightarrow s}\{\tau^{-\frac{n}{n+1}}(1+\tau^{\frac{1}{1+n}})^{-\alpha-1}\}\\
  =&\frac{\alpha}{1+n}\frac{(1+n)^{1+\alpha}}{(2\pi)^n\Gamma(1+\alpha)s^{\frac{1}{1+n}}}\\
     &\cdot G_{n+2,n+1}^{n+1,n+2}
  \bigg(\frac{1}{s}\bigg|\begin{matrix}
  \Delta\big(1,\frac{n}{1+n}\big)&, \Delta(1+n,-\alpha)\\
  \Delta(1+n,0)&
\end{matrix}\bigg).
\end{split}
\end{equation}
Here and in the following, we use the symbol $\mathcal{O}[s^\nu]:=C s^\nu$ with $C$ being a constant. Then basing on the representation of Meijer G-function as generalized hypergeometric functions (\ref{app_a_2}) and utilizing the definition (\ref{app_a_4}), we finally obtain the asymptotic behavior of $m_0$ for large time $t$ (small $s$)
\begin{equation}\label{1.2.7}
\begin{split}
   \hat{m}_0=&s^{-\frac{1}{1+n}}\bigg[s^{\frac{1}{1+n}}\bigg(\sum_{j=0}^{\infty} \mathcal{O}[s^j]\bigg)+s^{\frac{\alpha+n+1}{n+1}}\bigg(\sum_{j=0}^{\infty} \mathcal{O}[s^j]\bigg)\\
   &+s^{\frac{\alpha+n}{n+1}}\bigg(\sum_{j=0}^{\infty} \mathcal{O}[s^j]\bigg)+\ldots+s^{\frac{\alpha+1}{n+1}}\bigg(\sum_{j=0}^{\infty} \mathcal{O}[s^j]\bigg)\bigg] \\
\sim& 1+\mathcal{O}[s]+
\mathcal{O}[s^{\frac{\alpha}{1+n}}].
\end{split}
\end{equation}
On the other hand, $m_1$ can also be obtained from its asymptotic behavior
\begin{equation}\label{1.2.8}
\begin{split}
  & \hat{m}_1 \\
   &=\mathcal{L}_{\tau\rightarrow s}\bigg\{\tau^{\frac{2-n}{1+n}}\frac{\alpha}{1+n}\frac{1}{(1+\tau^{\frac{1}{1+n}})^{1+\alpha}}\bigg\} \\
      &=\frac{\alpha}{1+n}\frac{(1+n)^{1+\alpha}}{\Gamma(1+\alpha)s^{\frac{3}{1+n}}}\\
     & ~~~\cdot G_{n+2,n+1}^{n+1,n+2}
  \bigg(\frac{1}{s}\bigg|\begin{matrix}
  \Delta\big(1,\frac{n-2}{1+n}\big)&, \Delta(1+n,-\alpha)\\
  \Delta(1+n,0)&
\end{matrix}\bigg)\\
&=s^{-\frac{3}{n+1}}\bigg[s^{\frac{3}{1+n}}\bigg(\sum_{j=0}^{\infty} \mathcal{O}[s^j]\bigg)+s^{\frac{\alpha+n+1}{n+1}}\bigg(\sum_{j=0}^{\infty} \mathcal{O}[s^j]\bigg)\\
&~~~ +s^{\frac{\alpha+n}{n+1}}\bigg(\sum_{j=0}^{\infty} \mathcal{O}[s^j]\bigg)+\ldots+s^{\frac{\alpha+1}{n+1}}\bigg(\sum_{j=0}^{\infty} \mathcal{O}[s^j]\bigg)\bigg]\\
&~~~   \sim \sum_{j=0}^{\infty} \mathcal{O}[s^j]+s^{\frac{\alpha-2}{1+n}}\bigg(\sum_{j=0}^{\infty}\mathcal{O}[s^j]\bigg).
\end{split}
\end{equation}
Before turning to calculate $M_1$, we need to first calculate the integral 
\begin{equation}\label{1.2.9}
\begin{split}
   \int_{\tau}^{\infty}\lambda(f^{-1}(\eta))f^{-1}(\eta)'d\eta=&1-\int_{0}^{\tau}
  \lambda(f^{-1}(\eta))f^{-1}(\eta)'d\eta \\
      =&(1+\tau^{\frac{1}{1+n}})^{-\alpha}.
\end{split}
\end{equation}
Then
\begin{equation}\label{1.2.10}
\begin{split}
     & \hat{M}_1
     \\
     &=\mathcal{L}_{\tau\rightarrow s}\{\tau^{\frac{2}{1+n}}(1+\tau^{\frac{1}{1+n}})^{-\alpha}\}\\
    & =\frac{(n+1)\alpha}{\Gamma(\alpha)s^{\frac{n+3}{n+1}}}\\
     & ~~~ \cdot G_{n+2,n+1}^{n+1,n+2}
  \bigg(\frac{1}{s}\bigg|\begin{matrix}
  \Delta\big(1,\frac{-2}{1+n}\big)&, \Delta(1+n,1-\alpha)\\
  \Delta(1+n,0)&
\end{matrix}\bigg)\\
&=s^{-\frac{n+3}{n+1}}\bigg[s^{\frac{n+3}{1+n}}\bigg(\sum_{j=0}^{\infty} \mathcal{O}[s^j]\bigg)+s^{\frac{\alpha+n}{n+1}}\bigg(\sum_{j=0}^{\infty} \mathcal{O}[s^j]\bigg)\\
&~~~ +s^{\frac{\alpha+n-1}{n+1}}\bigg(\sum_{j=0}^{\infty} \mathcal{O}[s^j]\bigg)
  +\ldots+s^{\frac{\alpha}{n+1}}\bigg(\sum_{j=0}^{\infty} \mathcal{O}[s^j]\bigg)\bigg]\\
& \sim \sum_{j=0}^{\infty} \mathcal{O}[s^j]+s^{\frac{\alpha-n-3}{1+n}}\bigg(\sum_{j=0}^{\infty}\mathcal{O}[s^j]\bigg).
\end{split}
\end{equation}
Finally, we get the asymptotic behaviour of $\big<x^2(t)\big>$ with its Laplace transform
\begin{equation}\label{1.2.11}
  \big<x^2(s)\big>=\frac{m_1+s M_1}{s(1-m_0)}\sim\frac{1+ \mathcal{O}[s^{\frac{\alpha-2}{n+1}}]}{\mathcal{O}[s^2]+ \mathcal{O}[s^{\frac{\alpha+n+1}{n+1}}]}.
\end{equation}
Equation (\ref{1.2.11}) is the asymptotic behaviour of MSD with the velocity $v(\rho)=\frac{1}{\rho^n}$ and power law walking length distribution.
When $n=1$, then from (\ref{1.2.11}) there exists
\begin{equation}\label{1.2.13}
  \big<x^2(s)\big>\sim\frac{C_1+C_2 s^{\frac{\alpha-2}{2}}}{C_3 s^2+C_4 s^{\frac{\alpha+2}{2}}},
\end{equation}
where after some calculations $C_1=C_3=\frac{2}{(\alpha-1)(\alpha-2)}$ and $C_2=C_4=\Gamma(1-\frac{\alpha}{2})$. Then we can conclude that $\big<x^2(s)\big>\sim\frac{1}{s^2}$, i.e., $\big<x^2(t)\big>\sim t$, which also indicates that the choice of walking length distribution doesn't influence the MSD.
When $n>1$ and $\alpha>2$, we have
\begin{equation}\label{1.2.14}
  \big<x^2(s)\big>\sim \frac{1}{\mathcal{O}[s^2]+ \mathcal{O}[s^{\frac{\alpha}{1+n}+1}]},
\end{equation}
implying that
\begin{equation}\label{1.2.15}
  \big<x^2(t)\big>\sim
  \begin{cases}
  t, & \mbox{if $\alpha>n+1$,}\\
  t^{\frac{\alpha}{1+n}}, & \mbox{if $2<\alpha<n+1$},
  \end{cases}
\end{equation}
which indicates this kind of L\'{e}vy walk becomes subdiffusion when $2<\alpha<n+1$. On the other hand, if $0<\alpha<2$, then  $\frac{\alpha+n+1}{1+n}<2$
and there exists
\begin{equation}\label{1.2.16}
\big<x^2(s)\big>\sim \frac{\mathcal{O}[s^{\frac{\alpha-2}{1+n}}]}{\mathcal{O} [s^2] +\mathcal{O}[s^{\frac{\alpha+n+1}{1+n}}]}\sim s^{\frac{\alpha-2}{1+n}-\frac{\alpha+n+1}{1+n}}=s^{\frac{-3-n}{1+n}}.
\end{equation}
\begin{figure}
  \centering
  \includegraphics[width=8cm]{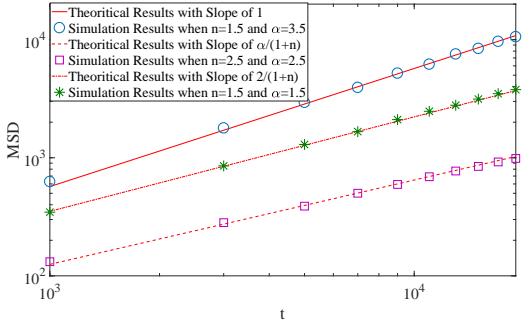}
  \caption{Numerical simulations of MSD of L\'{e}vy walk under the conditions of $n>1$, $\alpha>2$, and $0<\alpha<2$ by sampling over $10^4$ realizations (in log-log scale). The circles, stars, and squares represent the simulation results of symmetric L\'{e}vy walk with velocities $v(\rho)=1/\rho^{1+n}$ and walking length distributions $\lambda(\rho)=\frac{\alpha}{(1+\rho)^{1+\alpha}}$, where $n=1.5$, $\alpha=3.5$ (for circles), $n=1.5$, $\alpha=1.5$ (for stars) and $n=2.5$, $\alpha=2.5$ (for squares), respectively. The real, dotted, and dashed lines represent the corresponding theoretical results with the slope of $1$, $2/(n+1)=2/2.5$, and $\alpha/(1+n)=2.5/3.5$.
  }\label{fig_2}
\end{figure}
That is
\begin{equation}\label{msd_ex2_alpha<2}
  \big<x^2(t)\big>\sim t^{\frac{2}{1+n}},
\end{equation}
indicating a subdiffusion for $n>1$. Here in order to let the indexes of G-function used through the calculations make sense, $n$ must be an integer. However according to the simulation  results shown in Fig. \ref{fig_2}, one can conclude that the results shown in (\ref{1.2.15}) and (\ref{msd_ex2_alpha<2}) can be extended the domain of $n$ into real number that is bigger than 1. From (\ref{msd_ex2_alpha<2}), one can note that $\alpha$ has no influence on MSD when $0<\alpha<2$ and $n>1$. Besides it can be noted that when $n\geq 1$ and the velocity $v(\rho)=\frac{1}{\rho^n}$, the L\'{e}vy walk always show subdiffusion and normal diffusion. However when $0<n<1$, from (\ref{1.2.11}) it can be predicted that
\begin{equation}\label{1.2.17}
  \big<x^2(t)\big>\sim
  \begin{cases}
    t & \mbox{if $\alpha>2$,} \\
    t^{1+\frac{2-\alpha}{n+1}} & \mbox{if $n+1<\alpha<2$,}\\
    t^{\frac{2}{1+n}} & \mbox{if $\alpha<n+1$},
  \end{cases}
\end{equation}
implying that when $0<n<1$, the L\'{e}vy walk will always show superdiffusion or normal diffusion, which is verified by the numerical simulations given in Fig. \ref{fig_3}. All the results are summarized in Tab. \ref{tab.1}.


\begin{table}
\begin{tabular}{c|c|c|c}
\toprule
region of $n$ & region of $\alpha$ & MSD & category of diffusion\\
\colrule
      \multirow{3}{*}{$0<n<1$} & $0<\alpha<n+1$ & $\sim t^{\frac{2}{1+n}}$ & \multirow{2}{*}{superdiffusion}\\
      \cline{2-3}
      \multirow{3}{*}{} & $n+1<\alpha<2$ & $\sim t^{1+\frac{2-\alpha}{n+1}}$ & \multirow{2}{*}{}\\
      \cline{2-4}
      \multirow{3}{*}{} & $\alpha>2$ & $\sim t$ & \multirow{3}{*}{normal diffusion} \\
      \cline{1-3}
       $n=1$ & all $\alpha>0$ &{$=t$}&  \multirow{3}{*}{}\\
       \cline{1-3}
     \multirow{3}{*}{$1<n$} & $n+1<\alpha$ & $\sim t$ & \multirow{3}{*}{} \\
     \cline{2-4}
      \multirow{3}{*}{} & $2<\alpha<n+1$ & $\sim t^{\frac{\alpha}{1+n}}$ & \multirow{2}{*}{subdiffusion} \\
      \cline{2-3}
      \multirow{3}{*}{} & $0<\alpha<2$ & $\sim t^{\frac{2}{1+n}}$ & \multirow{2}{*}{} \\
\botrule
\end{tabular}
\caption{
MSDs classified by $n$ and $\alpha$.
}
\label{tab.1}
\end{table}

%
%
%
%
%
%
%
%
%
%
%
%
%
%
%
%
\subsubsection{Being $\rho/(\exp(\rho)-1)$ for velocity function $v(\rho)$
}
\begin{figure}
  \centering
  \includegraphics[width=8cm]{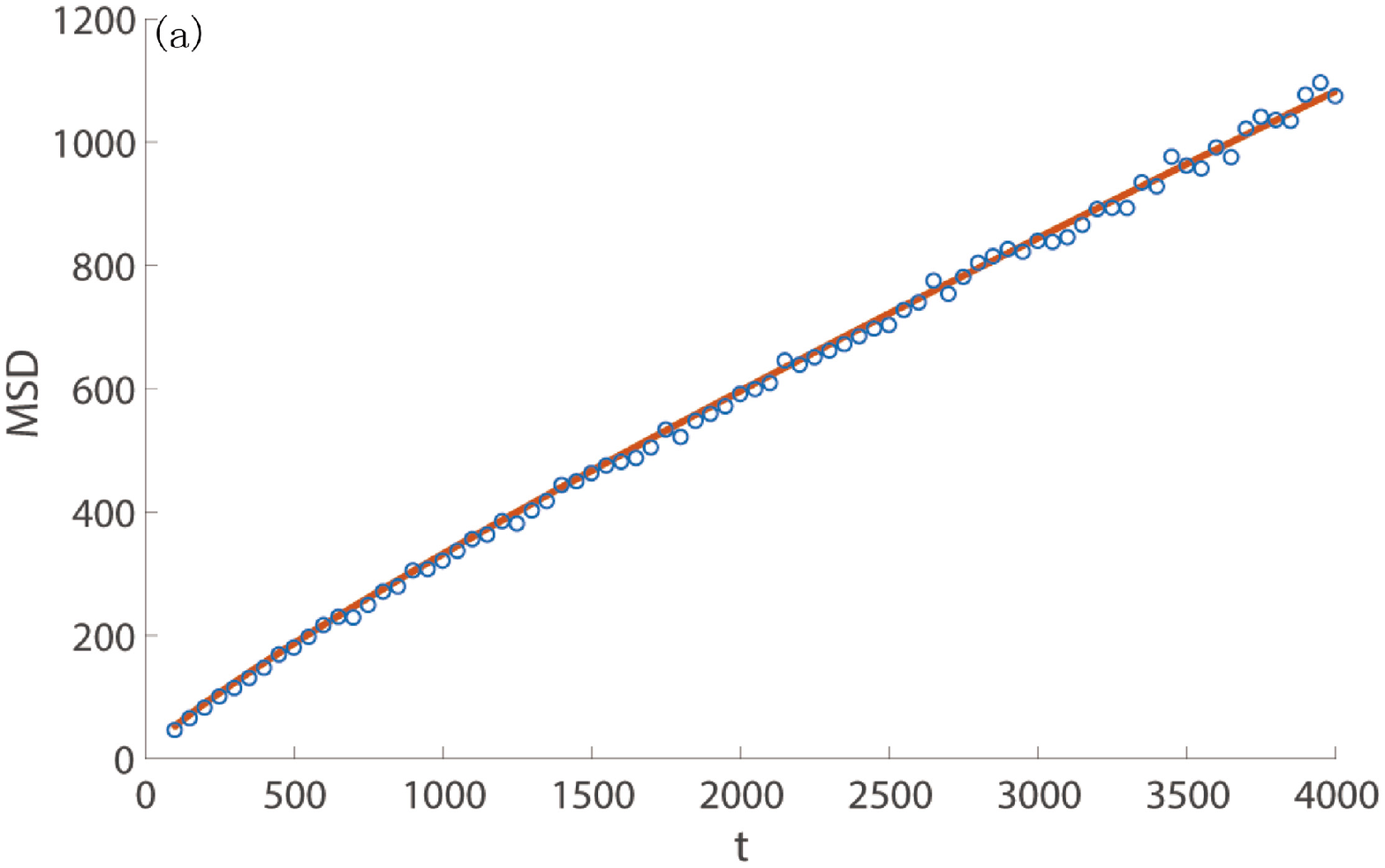}
  \includegraphics[width=8cm]{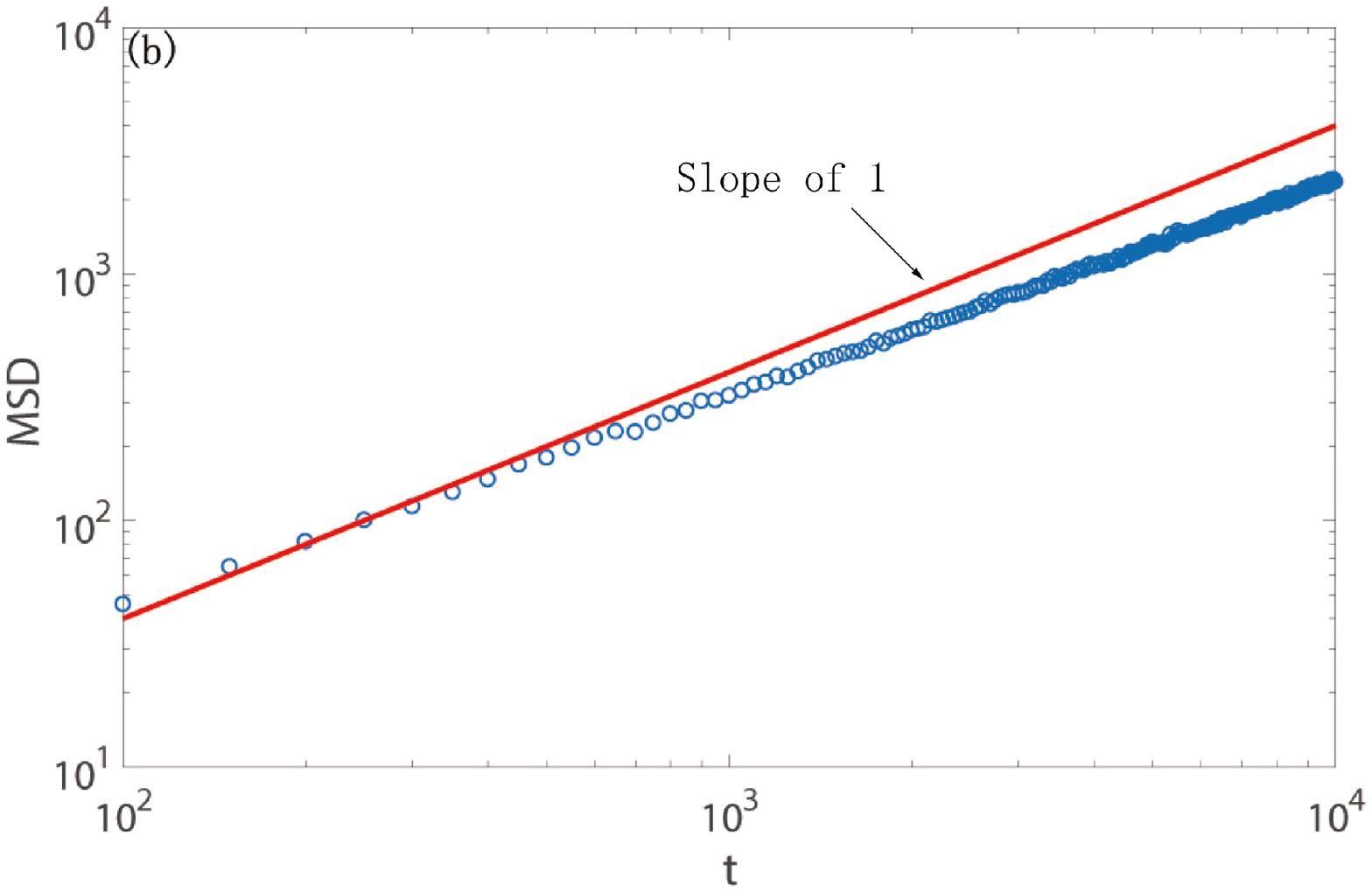}
  \caption{Numerical simulations of MSD of L\'evy walk with the velocity $v(\rho)=\frac{\rho}{\exp(\rho)-1}$ by sampling over $10^4$ realizations. The walking length distribution follows $\lambda(\rho)=\exp(-\rho)$. In Figures (a) and (b), the circles represent the simulation results. The (red) solid line in Figure (a) is the theoretical result obtained from the numerical inverse Laplace transform of (\ref{asymp_msd_exp_vrho}). 
  Figure (b) tries to compare the MSD of this L\'evy walk with normal diffusion, the slope of the MSD of which is one.
  }\label{fig_exp_vrho}
\end{figure}

We further consider the velocity function $v(\rho)=\frac{\rho}{\exp(\rho)-1}$, which indicates the longer distance a particle walks, the longer time it may take, and such walking duration increases exponentially with $\rho$. Then one can easily obtain $f(\rho)=\exp(\rho)-1$, i.e., $\rho=f^{-1}(\tau)=\ln (1+\tau)$.
As for $\lambda(\rho)$, we first take it to be $\mu\exp(-\mu\rho)$ with $\mu>0$.
In order to obtain MSD by utilizing (\ref{MSD_vrho}), we still need to calculate $\hat{m}_0$, $\hat{m}_1$, and $\hat{M}_1$ from (\ref{1.2.19}). After calculations, there is
\begin{equation}\label{m0_exp_vrho}
  \hat{m}_0(s)=\mu \exp(s) E_{1+\mu}(s),
\end{equation}
where $E_{\mu}(s)$ represents the exponential integral function; and it satisfies the asymptotic behavior
\begin{equation}
  \hat{m}_0(s)\sim \begin{cases}
                     \exp(s)(1+(\gamma-1+\ln(s))s) & \mbox{if $\mu=1$},\\
                     \mu\exp(s)(s^\mu \Gamma(-\mu)+\frac{1}{\mu}+\frac{s}{1-\mu}) & \mbox{otherwise},
                   \end{cases}
\end{equation}
where $\gamma$ is Euler's constant with approximate value $0.577216$. And
\begin{equation}\label{m1_M_exp_vrho}
  \begin{split}
     \hat{m}_1 & =2 \mu \exp(s) G_{3,4}^{4,0}
  \bigg(s\bigg|\begin{matrix}
  &\mu+1,&\mu+1,&\mu+1\\
  &0,&\mu,&\mu,&\mu
\end{matrix}\bigg), \\
    \hat{M}_1 & =2 \exp(s) G_{3,4}^{4,0}
  \bigg(s\bigg|\begin{matrix}
  &\mu,&\mu,&\mu\\
  &0,&\mu-1,&\mu-1,&\mu-1
\end{matrix}\bigg).
  \end{split}
\end{equation}
When $\mu=1$, for big $t$ there exists the asymptotic behavior in Laplace domain
\begin{equation}\label{asymp_msd_exp_vrho}
  \big<x^2(s)\big>\sim-\frac{2 \exp(s)}{s^2(\gamma+\ln(s))}\sim-\frac{2}{s^2 \ln(s)}.
\end{equation}
When $s$ tends to zero, for any given positive $\epsilon$, $s^2<-s^2\ln(s)<s^{2-\epsilon}$. So, the MSD grows slower than $t$ but faster than $t^{1-\epsilon}$ when $t$ goes to infinity; see Fig. \ref{fig_exp_vrho}.

\subsection{Some notes and discussions}

In the previous subsections, we mainly focus on the L\'evy walk with its velocity depending on the walking length of each step. Further, it's natural to ask what happens if the velocity is a function of the walking duration $\tau$ of each step. It can be shown that essentially they are equivalent. In fact, following the derivation of (\ref{1.1.6}), when $v=\mathrm{v}(\tau)$, the PDF $P(x,t)$ in the Fourier-Laplace space can also be obtained as
\begin{equation}\label{1.3.1}
  \hat{\bar{P}}(k,s)=\frac{\mathcal{L}_{\tau\rightarrow s}\{\cos({\mathrm{v}}(\tau)\tau k)\int_{\tau}^{\infty}\phi(\eta)d\eta\}}{1-\mathcal{L}_{\tau\rightarrow s}\{\cos(\mathrm{v}(\tau)\tau k)\phi(\tau)\}},
\end{equation}
where $\phi(\tau)$ is the PDF of walking time for each step of moving. Considering the facts $\rho=\mathrm{v}(\tau)\tau$ and $f(\rho)\equiv \tau=\frac{\rho}{v(\rho)}$, we obtain $\rho=f^{-1}(\tau)$, and then $\mathrm{v}(\tau)=\frac{f^{-1}(\tau)}{\tau}$. That is to say, we can transfer the $v(\rho)$ and $\lambda(\rho)$ equivalently into $\mathrm{v}(\tau)$ and $\phi(\tau)=\lambda(f^{-1}(\tau))f^{-1}(\tau)'$.

\begin{figure}
  \centering
  \includegraphics[width=8cm]{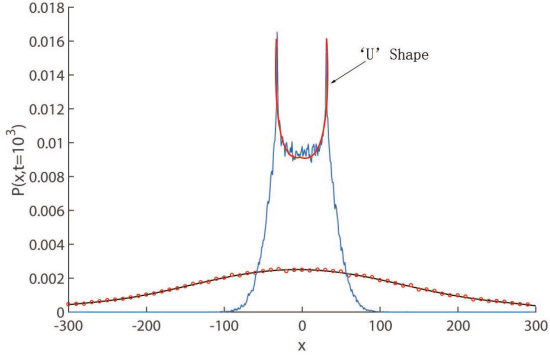}
  \caption{Comparison of the PDFs of ordinary L\'{e}vy walk and L\'{e}vy walk with the velocity $v(\rho)=1/\rho^n$ by sampling over $10^5$ realizations. The flatter one is symmetric L\'{e}vy walk (at time $t=10^3$) with $v=1$ and walking length distribution $\lambda(\rho)=\alpha/(1+\rho)^{1+\alpha}$, where $\alpha=1.5$. All the parameters of the taller one with a `U' shape are the same as the ones of the flatter one except $v=1/\rho$.
   }\label{fig_PDF_1}
\end{figure}

\begin{figure}
  \centering
  \includegraphics[width=8cm]{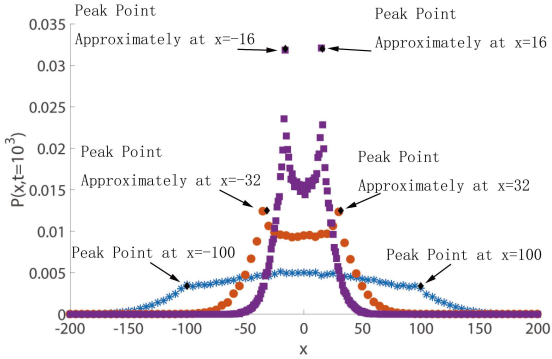}
  \caption{(Colored online) Simulations of PDFs of  symmetric L\'{e}vy walks with the velocity $v=1/\rho^{n}$ by sampling over $10^5$ realizations at time $t=10^3$. All the walking length distributions are the same, being $\lambda(\rho)=\alpha/(1+\rho)^\alpha$ with $\alpha=1.5$. The velocities are chosen as $v=1/\rho^n$ respectively with the parameter $n=0.5$ (blue stars), $n=1$ (orange circles),  and $n=1.5$ (purple squares). 
  }\label{fig_PDF_2}
\end{figure}

By comparing the PDFs, here we would like to make some notes on the major differences between the L\'{e}vy walk with constant velocity and with the velocity depending on the  moving length $\rho$ of each step. The dotted line in Fig. \ref{fig_PDF_1} is for the L\'{e}vy walk with velocity $v=1$, while the taller one is for the L\'{e}vy walk with velocity $v(\rho)=1/\rho$; their walking length density is $\lambda(\rho)=\alpha/(1+\rho)^{1+\alpha}$ with $\alpha=1.5$. It can be seen that the main feature of the PDF for nonconstant $v(\rho)$ is the appearance of `U' shape; more details, including the dependence of the shape on the parameters, will be discussed in the next paragraph. Now, we first turn to discuss the position of the high peak of the PDF.  For the L\'{e}vy walk (starting from the origin) with velocity $v(\rho)=1/\rho^n$ and $n>0$, or equivalently $\mathrm{v}(\tau)=f^{-1}(\tau)/\tau=\tau^{-n/(n+1)}$, the position of the particle is $\pm \mathrm{v}(t)t=\pm t^{1/(n+1)}$ if the particle does not finish its first step till time $t$, otherwise the farthest position that the particle can reach is bigger than $t^{1/(n+1)}$ since $t_1^{1/(n+1)}+t_2^{1/(n+1)}>(t_1+t_2)^{1/(n+1)}$. From the simulations shown in Fig. \ref{fig_PDF_2}, one can note that $\pm t^{1/(n+1)}$ are the positions of the high peaks of the PDF if it has.

%

\begin{figure}
  \centering
  \includegraphics[width=8cm]{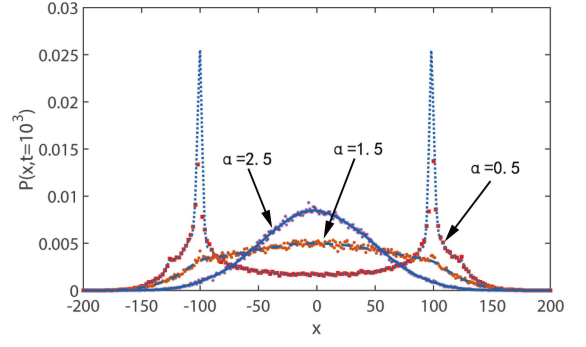}
  \caption{(Colored online) Simulations of PDFs of symmetric L\'{e}vy walks with velocity $v=1/\rho^n$ at time $t=10^3$, by choosing different walking length distribution and sampling over $10^5$ realizations. The walking length distributions are $\lambda(\rho)=\alpha/(1+\rho)^{1+\alpha}$ with the parameters taken as $n=0.5$, $t=10^3$, $\alpha=0.5$ (dotted line with red squares), $\alpha=1.5$ (dashed line with orange circles), and $\alpha=2.5$ (real line with purple stars).}\label{fig_PDF_3}
\end{figure}

\begin{figure}
  \centering
  \includegraphics[width=8cm]{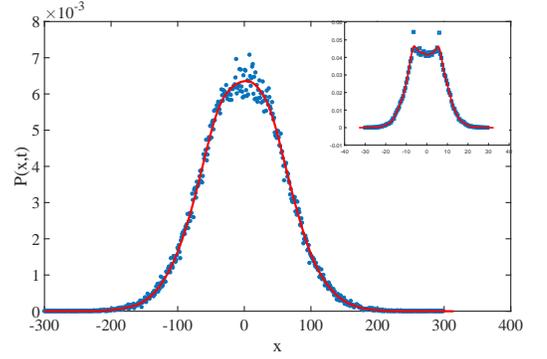}
  \caption{Simulations of PDFs of symmetric L\'{e}vy walks with velocity $v=1/\rho^{1.5}$ at different times by sampling over $10^5$ realizations. The walking length distributions are the same, $\lambda(\rho)=\alpha/(1+\rho)^{1+\alpha}, \alpha=1.5$; while the observation times are $t=10^4$ and $t=10^2$ (the inner picture).}\label{fig_PDF_4}
\end{figure}

According to Fig. \ref{fig_PDF_2}, it also turns out that the `U' shape doesn't always appear. More precisely, Fig. \ref{fig_PDF_3} shows that the shapes of PDF relate to the choice of $\alpha$. For $0<\alpha<1$, the `U' shape always exists; for $1<\alpha<2$ and $n>1$, the `U' shape can also be observed (not shown in Fig. \ref{fig_PDF_3}); for $1<\alpha<2$ and $0<n<1$, the `U' shape disappears, however there are still 2 sharp points at $x=\pm t^{1/(n+1)}$, respectively; for $\alpha>2$ and $0<n<1$, the PDF turns out to be smooth. Besides from Fig. \ref{fig_PDF_4}, it can be observed that if taking $\alpha>2$ and $n>2$, when $t$  is large enough, the `U' shape gradually disappears.
The above phenomena may be considered as the effect of heavy tail. When $0<\alpha<1$, due to the dominant effect of heavy tail, there is a big probability that the particle hasn't finished its  first step, and it will cause peaks around the points $\pm \mathrm{v}(\tau)\tau$. For the other cases, the tail becomes less heavy, so it forms a competition between $\alpha$ and $n$.

\section{L\'{e}vy walk with velocity depending on the current position}\label{sec3}
%
%

In this section, we discuss the L\'{e}vy walk with velocity depending on the current position, i.e., $v=v(x)$. As a concrete example, we take $v(x)=v_0-\frac{c}{\tau}x$, where $c$ is a positive constant, and $v_0$ is a non-zero constant. It seems that the integral transform method does not work for this model, and we have to resort to Hermite polynomial expansion approach. When $v_0=0$ or $c=1$, the model will be trivial.  If $v_0=0$, the particle will soon be attracted to the position $x=0$, meaning $P(x,t)=\delta(x)$ after a long time. If $c=1$, then $\delta(\rho+(\frac{c}{\tau}(x-\rho)-v_0)\tau)= \delta(x-v_0 \tau)$. The governing equation of $q(x,t)$ becomes
\begin{equation}\label{2.2.1}
  \begin{split}
     & q(x,t) \\
      &     = \int_{-\infty}^{\infty}\int_{0}^{t}q(x-\rho,t-\tau) \phi(\tau) \\
     &~~~ \cdot\delta\big(x-v_0 \tau\big)d\tau d\rho
        +P_0(x)\delta(t).
  \end{split}
\end{equation}

\begin{figure}
  \centering
  \includegraphics[width=8cm]{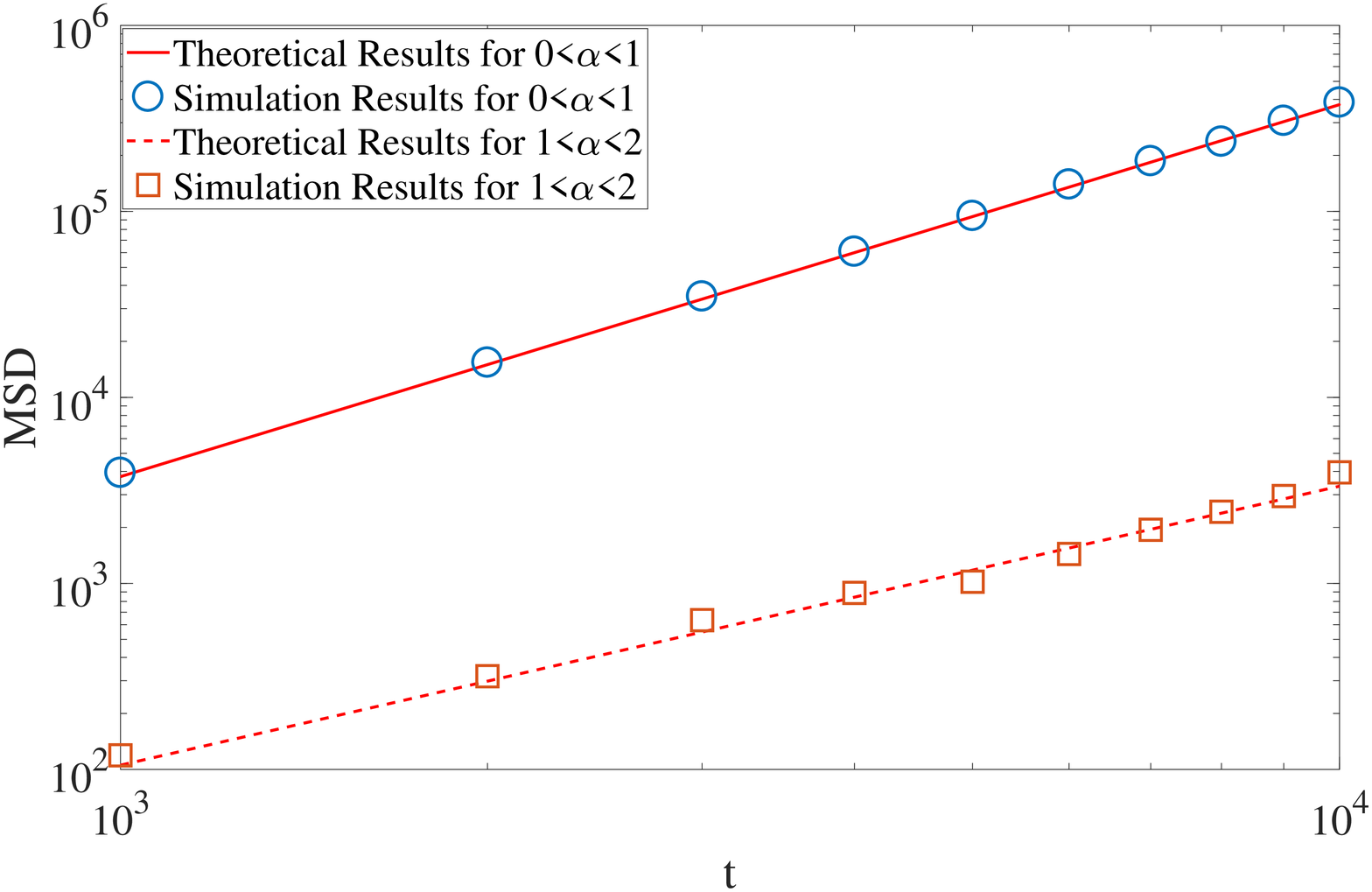}
  \includegraphics[width=8cm]{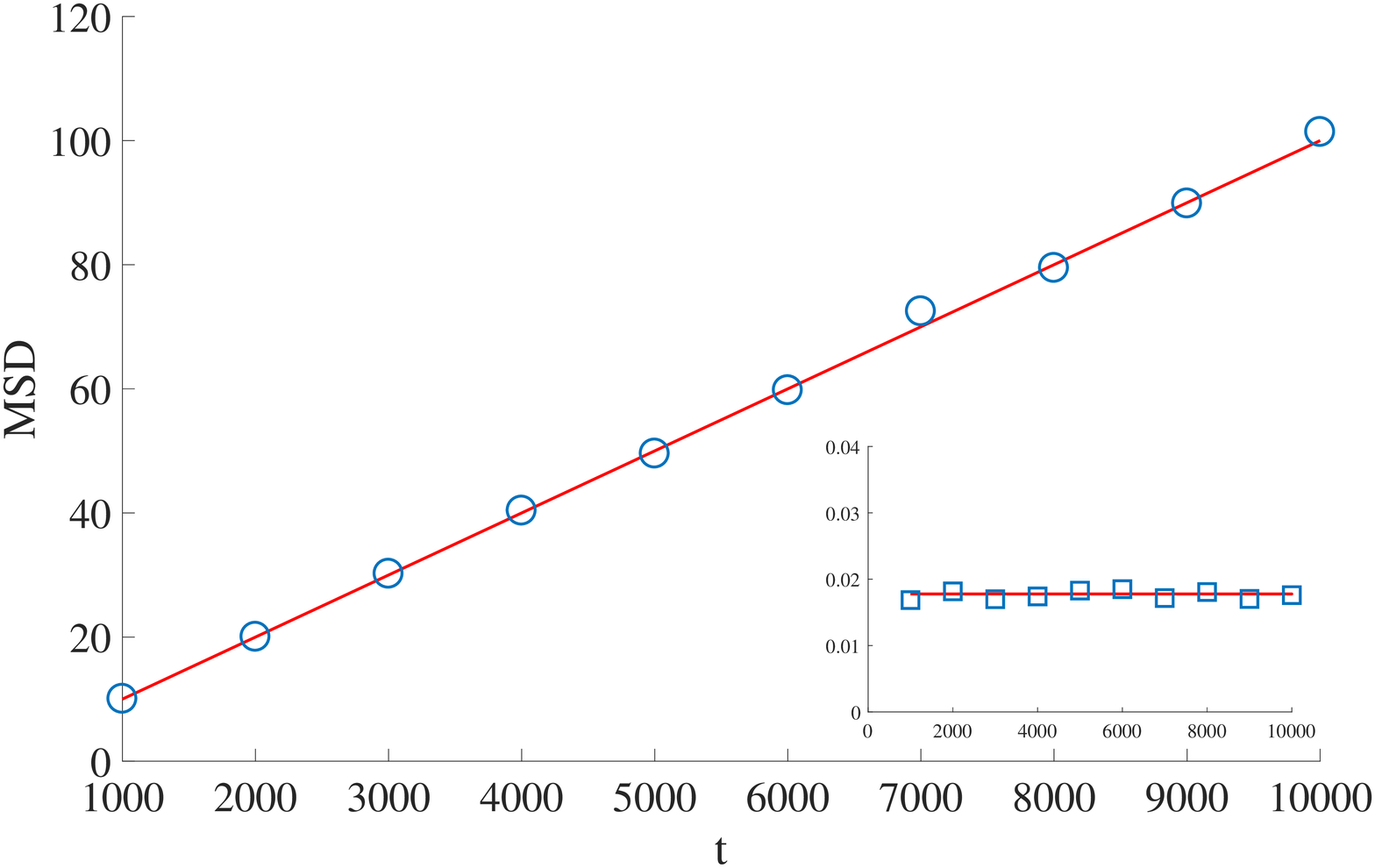}
  \caption{Simulations of MSD of L\'{e}vy walk with velocity $v(x)=v_0-\frac{c}{\tau}x$. Each dot is obtained by averaging over $10^4$ realizations. The above figure shows the simulations of MSD of L\'{e}vy walk with $\alpha=0.5$, $c=1.5$ (circles) and $\alpha=1.5$, $c=1.5$ (squares). The lower one illustrates the L\'{e}vy walk with $\phi(\tau)=\exp(-\tau)$, and $c=2$ (circles), $c=1.5$ (squares in the inner figure). 
  }\label{fig_4}
\end{figure}

In the following, we assume that $c\neq1$ and $v_0\neq 0$.  After some calculations, there exists
\begin{equation}\label{2.2.2}
  q(x,t)=\frac{1}{|c-1|}\int_{0}^{t}q\bigg(\frac{v_0\tau-x}{c-1},t-\tau\bigg)\phi(\tau)d\tau+P_0(x)\delta(t).
\end{equation}
Here we still use Hermite polynomials in (\ref{2.1.3}) to approach $q(x,t)$. After multiplying $H_m(x)$ and integrating with respect to $x$, we have
\begin{equation}\label{2.2.3}
\begin{split}
&\sqrt{\pi}2^m m!T_m(t)\\
&=\sum_{n=0}^{\infty}\int_{0}^{t}\frac{1}{|c-1|}\int_{-\infty}^{\infty}H_n\bigg(\frac{v_0 \tau-x}{c-1}\bigg)T_n(t-\tau)H_m(x)dx \\
     &~~~~ \cdot \phi(\tau)d\tau+
    H_m(x_0)\delta(t).
\end{split}
\end{equation}
First note that
\begin{equation}\label{2.2.4}
\begin{split}
&\frac{1}{|c-1|}\int_{-\infty}^{\infty}H_n\left(\frac{v_0\tau-x}{c-1}\right)\exp\left(-\left(\frac{v_0\tau-x}{\tau-1}\right)^2\right)\\
&~~~~\cdot H_m(x)dx\\
&=\int_{-\infty}^{\infty}H_n(y)\exp(-y^2)H_m(v_0\tau-(c-1)y)dy.
\end{split}
\end{equation}
Then basing on the properties of Hermite polynomials (\ref{App_B_3}), (\ref{App_B_5}), and (\ref{App_B_6}),
the equation for $H_m(v_0 \tau-(c-1)y)$ can be obtained as
\begin{equation}\label{2.2.5}
  \begin{split}
     & H_m(v_0 \tau-(c-1)y)  \\
       &=\sum_{k=0}^{m}\frac{m!}{k!(m-k)!}H_k(-(c-1)y)(2v_0\tau)^{m-k}\\
       &=\sum_{k=0}^{m}\frac{m!}{(m-k)!}(2v_0\tau)^{m-k}\sum_{i=0}^{\lfloor\frac{k}{2}\rfloor}\frac{(1-c)^{k-2i}(c^2-2c)^i}{(k-2i)!i!}\\
       &~~~~\cdot H_{k-2i}(x).
  \end{split}
\end{equation}
Substituting (\ref{2.2.5}) into (\ref{2.2.3}) and taking Laplace transform w.r.t. $t$ lead to the recurrence relation  
\begin{equation}\label{2.2.6}
  \begin{split}
     &\sqrt{\pi}2^m m!\hat{T}_m(s)  \\ &=\sum_{k=0}^{m}\sum_{i=0}^{\lfloor\frac{k}{2}\rfloor}\frac{m! 2^{m-2 i\sqrt{\pi}}}{(m-k)!i!}\hat{T}_{k-2i}(s) \\
       &~~~~\cdot \mathcal{L}_{\tau\rightarrow s}\{(v_0\tau)^{m-k}(1-c)^{k-2i}(c^2-2c)^{i}\phi(\tau)\}\\
       &~~~~+H_m(x_0).
  \end{split}
\end{equation}
From (\ref{2.2.6}), we can obtain
\begin{eqnarray}
 \hat{T}_0(s) &=& \frac{1}{\sqrt{\pi}(1-\hat{\phi}(s))}, \\
  \hat{T}_1(s) &=& \frac{-x_0+x_0\hat{\phi}(s)+v_0\hat{\phi}'(s)}{\sqrt{\pi}(\hat{\phi}(s)-1)(1+(c-1)\hat{\phi}(s))},
\end{eqnarray}
and
\begin{equation}\label{2.2.7}
  \begin{split}
       \sqrt{\pi}2^3\hat{T}_2(s) =&2^2\sqrt{\pi}v_0^2\hat{\phi}''(s)\hat{T}_0(s)-2^3\sqrt{\pi}v_0(1-c)\\
       & \cdot\hat{\phi}'(s)\hat{T}_1(s) +2^3\sqrt{\pi}(1-c)^2\hat{\phi}(s)\hat{T}_2(s)\\
       &+2\sqrt{\pi}(c^2-2c)\hat{\phi}(s)\hat{T}_0(s)
       +H_2(x_0).
  \end{split}
\end{equation}
Then we consider the PDF of finding the particle at position $x$ at time $t$, which can be obtained as 
\begin{equation}\label{2.2.7}
\begin{split}
    P(x,t)=&\int_{-\infty}^{\infty}dy \int_{0}^{t}q(x-y,t-\tau)\Psi(\tau)\\
    &\cdot\delta\Big(y+\Big(\frac{c}{\tau}(x-y)-v_0\Big)\tau\Big)d\tau.
\end{split}
\end{equation}
Rewrite $P(x,t)$ as the form of (\ref{2.1.4}). After taking Laplace transform w.r.t.  $t$, we have
\begin{equation}\label{2.2.8}
\begin{split}
   & \hat{\tilde{T}}_m(s) 
   \\
   &=\sum_{k=0}^{m}\sum_{i=0}^{\lfloor\frac{k}{2}\rfloor}\frac{2^{-2i}}{(m-k)!i!}\hat{T}_{k-2i}(s)  \\
     & ~~~ \cdot\mathcal{L}_{\tau\rightarrow s}\{(v_0 \tau)^{m-k}(1-c)^{k-2i}(c^2-2c)^{i}\Psi(\tau)\}.
\end{split}
\end{equation}
Then
\begin{equation}\label{2.2.9}
  \begin{split}
     \hat{\tilde{T}}_0(s) & = \hat{\Psi}(s)\hat{T}_0(s)=\frac{1}{\sqrt{\pi}s}, \\
      \hat{\tilde{T}}_1(s) & = -v_0\hat{\Psi}'(s)\hat{T}_0(s)+(1-c)\hat{\Psi}(s)\hat{T}_1(s),\\
       \hat{\tilde{T}}_2(s)&=\frac{1}{2}v_0^2\hat{\Psi}''(s)\hat{T}_0(s)-v_0(1-c)\hat{\Psi}'(s)\hat{T}_1(s)\\
       &~~~~+(1-c)^2\hat{\Psi}(s)\hat{T}_2(s)
  +\frac{1}{4}(c^2-2c)\hat{\Psi}(s)\hat{T}_0(s).
  \end{split}
\end{equation}
According to (\ref{2.1.4}), there exists 
\begin{equation}\label{2.2.11}
 \begin{split}
    \big<x(s)\big> &=\sqrt{\pi}\hat{\tilde{T}}_1(s) \\
     \big<x^2(s)\big> & =\frac{\sqrt{\pi}}{2}\hat{\tilde{T}}_0(s)+2\sqrt{\pi}\hat{\tilde{T}}_2(s) \\
      & =\frac{1}{2s}+2\sqrt{\pi}\hat{\tilde{T}}_2(s).
 \end{split}
\end{equation}
In the following, we consider different kinds of flight time distributions $\phi(\tau)$. First we investigate the power-law flight time as the form shown in (\ref{L_powerlaw}). For $0<\alpha<1$, $\hat{\phi}(s)\sim 1-\Gamma(1-\alpha)s^\alpha$. After some calculations, we can obtain
\begin{eqnarray}
  \big<x(s)\big>&\sim&\frac{(1-\alpha)v_0}{s^2},\\
  \big<x^2(s)\big>&\sim&\frac{(2-\alpha)(1-\alpha)v_0^2}{s^3}.
\end{eqnarray}
Thus the corresponding inverse Laplace transforms are
\begin{eqnarray}
  \big<x(t)\big>&\sim&(1-\alpha)v_0t, \\
  \big<x^2(t)\big>&\sim&\frac{(2-\alpha)(1-\alpha)v_0^2}{2}t^2.
\end{eqnarray}
For $1<\alpha<2$, $\hat{\phi}(s)\sim1-\frac{1}{\alpha-1}s-\Gamma(1-\alpha)s^\alpha$, and
\begin{eqnarray}
  \big<x(t)\big> &=& \frac{(\alpha-1)v_0}{2-\alpha}t^{2-\alpha},\\
  \big<x^2(t)\big> &=& \frac{(\alpha-1)v_0^2}{3-\alpha}t^{3-\alpha}.
\end{eqnarray}
For the exponential distribution, we simply consider the Laplace transform with the form of $\hat{\phi}(s)=\frac{1}{1+s}$. Then

\begin{eqnarray}
  \big<x(t)\big> &=& \frac{v_0}{c}, \label{2.2.13}\\
  \big<x^2(t)\big> &=& \frac{2v_0^2}{(2-c)c^2}~~~~\mathrm{for}~~ c<2,\label{2.2.14}\\
  \big<x^2(t)\big> &=& v_0^2 t~~~~~~~~~~~~\mathrm{for}~~c=2,\label{2.2.15}
\end{eqnarray}
According to (\ref{2.2.13}) and (\ref{2.2.14}), it turns out that the particle has a localization. The results above are verified by the numerical simulations shown in Fig. \ref{fig_4}.

%
%
%

\section{Conclusion}\label{sec5}

This paper discusses the L\'{e}vy walk with velocity depending on walking length or walking time of each step, and with velocity being a function of current position. When doing the dynamical analyses of the L\'{e}vy walk, we introduce the Hermite polynomial expansion approach, which can be effectively used to analyze the time-space coupled or nonlinear models. This approach is an important complement to integral transform method. Both the integral transform method and Hermite polynomial expansion approach work for the L\'{e}vy walk with velocity depending on walking length or walking time of each step. One of the striking results is that when $v(\rho)=1/\rho$ the process will always show a normal diffusion, no matter what kind of walking length distribution is. By numerical simulations, the rich structure information of the PDF is uncovered.  As for the L\'{e}vy walk with velocity being a function of current position, we use the Hermite polynomial expansion approach to do the analysis with also some interesting phenomena uncovered. This kind of orthogonal polynomial approaches will be further developed in the coming researches.

\section*{Acknowledgements}
This work was supported by the National Natural Science Foundation of China under grant no. 11671182, and the Fundamental Research Funds for the Central Universities under grant no. lzujbky-2018-ot03.
\begin{appendix}
\section{A brief introduction of Meijer G-function and generalized hypergeometric function}
Here we only briefly introduce the definition and some important properties of Meijer G-function. For more details, one can refer to \cite{prudnikov}.
The Meijier G-function of order $(m,n,p,q)$, where $0\leq m\leq q$ and $0\leq n\leq p$, is defined as
\begin{widetext}
\begin{equation}\label{app_a_1}
  G_{p,q}^{m,n}\bigg(z\bigg|\begin{matrix}
(a_p)  \\
(b_q)
\end{matrix}\bigg)\equiv G_{p,q}^{m,n}\bigg(z\bigg|\begin{matrix}
a_1,\ldots,a_p \\
b_1,\ldots,b_q
\end{matrix}\bigg)=\frac{1}{2\pi i}\int_{L}\Gamma\bigg[
\begin{matrix}
  b_1+s,\ldots,b_m+s,1-a_1-s,\ldots,1-a_n-s \\
  a_{n+1}+s,\ldots,a_{p}+s,1-b_{m+1}-s,\ldots,1-b_q-s
\end{matrix}
\bigg]z^{-s}ds;
\end{equation}
\end{widetext}
where $L$ is a contour defined in \cite{prudnikov}.
One of the most important properties of Meijer G-function is the representation of generalized hypergeometric functions
\begin{widetext}
\begin{equation}\label{app_a_2}
\begin{split}
&G_{p,q}^{m,n}\bigg(z\bigg|\begin{matrix}
(a_p)  \\
(b_q)
\end{matrix}\bigg)=\\
&=\sum_{k=1}^{n}\Gamma\bigg[
\begin{matrix}
  a_k-a_1,\ldots,a_k-a_{k-1},a_k-a_{k+1},\ldots,a_k-a_n,1+b_1-a_k,\ldots,1+b_m-a_k \\
  a_k-b_{m+1},\ldots,a_k-b_q,1+a_{n+1}-a_{k},\ldots,1+a_p-a_k
\end{matrix}
\bigg] z^{a_k-1} \\
&~~~\cdot {_qF_{p-1}}\bigg(\begin{matrix}
1+(b_q)-a_k \\
1+(a_p)^{*}-a_k
\end{matrix};\frac{(-1)^{q-m-n}}{z}\bigg)
\end{split}
\end{equation}
\end{widetext}
under the assumptions $p\geq q$; $a_j-a_k\neq0,\pm1,\pm2,\cdots$; $j\neq k$; $j,k=1,2,\cdots,n$, and some requests on the contour $L$. 
Here $(a_p)^{*}-a_k\equiv a_1-a_k,\cdots,a_{k-1}-a_k,a_{k+1}-a_k,\cdots,a_p-a_k$;
\begin{equation}\label{app_a_3}
  \Gamma\bigg[
\begin{matrix}
  a_1,\cdots,a_p \\
  b_1,\cdots,b_q
\end{matrix}
\bigg]=\frac{\prod_{k=1}^{p}\Gamma(a_k)}{\prod_{l=1}^{q}\Gamma(b_l)},
\end{equation}
and the function $_qF_{p-1}$ in (\ref{app_a_2}) represents the generalized hypergeometric function, which is defined as
\begin{equation}\label{app_a_4}
  _pF_q\bigg(\begin{matrix}
         a_1,\cdots,a_p \\
         b_1,\cdots,b_q
       \end{matrix};z\bigg)=\sum_{n=0}^{\infty}\frac{(a_1)_n\cdots (a_p)_n}{(b_1)_n\cdots (b_q)_n}\frac{z^n}{n!}
\end{equation}
with $(a)_0=1$ and $(a)_n=a(a+1)(a+2)\cdots (a+n-1)$ for $n\geq 1$.

\section{A brief introduction to Hermite polynomials and orthogonal polynomials}

For a system of polynomials $f_n(x)$ with degree $[f_n(x)]=n$, if there is a function $\omega(x)\,[\omega(x)\geq0]$ on the interval $a\leq x\leq b$ such that
\begin{equation}\label{App_B_1}
  \int_{a}^{b}f_n(x)f_m(x)\omega(x)dx=0,
\end{equation}
for $n\neq m$ with $n,m=0,1,2,\cdots$, then the system of polynomials $f_n(x)$ is called orthogonal on the interval $a\leq x\leq b$  w.r.t. the weight function $\omega(x)$ \cite{abra}.
Following different kinds of weight functions and/or intervals, there are the corresponding orthogonal polynomials \cite{prudnikov,abra}. One of the most important kind of orthogonal polynomials defined on $(-\infty,\infty)$ are Hermite polynomials with the weight function $\omega(x)=e^{-x^2}$.
It can also be calculated as
\begin{equation}\label{App_B_2}
  H_n(x)=(-1)^n e^{x^2}\frac{d^n}{dx^n}e^{-x^2}.
\end{equation}
Its orthogonality is given as
\begin{equation}\label{App_B_3}
\int_{-\infty}^{\infty}H_m(x)H_n(x)e^{-x^2}dx=\sqrt{\pi}2^n n! \delta_{n,m},
\end{equation}
where $\delta_{n,m}$ is the Kronecker delta function.
Here we only illustrate some of the important properties of Hermite polynomials. The values of Hermite polynomials at $0$ are very useful
\begin{equation}\label{App_B_4}
  H_n(0)=
  \begin{cases}
    0 & \mbox{if $n$ is odd},\\
    (-2)^{\frac{n}{2}}(n-1)!! & \mbox{if $n$ is even},
  \end{cases}
\end{equation}
which indicates the recursion relation $H_n(0)=-2(n-1)H_{n-2}(0)$. Besides the following two expressions are also important in this paper:
\begin{equation}\label{App_B_5}
  H_n(x+y)=\sum_{k=0}^{n}\bigg(
  \begin{matrix}
    n \\
    k
  \end{matrix}
  \bigg)H_k(x)(2y)^{n-k}
\end{equation}
and
\begin{equation}\label{App_B_6}
  H_n(\gamma x)=\sum_{i=0}^{\lfloor\frac{n}{2}\rfloor}\gamma^{n-2i}(\gamma^2-1)^i
  \bigg(
  \begin{matrix}
    n \\
    2i
  \end{matrix}
  \bigg)
  \frac{(2i)!}{i!}H_{n-2i}(x),
\end{equation}
where $\lfloor\frac{n}{2}\rfloor$ is the biggest integer smaller than $\frac{n}{2}$.

\end{appendix}

\bibliography{ref}

\begin{thebibliography}{21}%
\makeatletter
\providecommand \@ifxundefined [1]{%
 \@ifx{#1\undefined}
}%
\providecommand \@ifnum [1]{%
 \ifnum #1\expandafter \@firstoftwo
 \else \expandafter \@secondoftwo
 \fi
}%
\providecommand \@ifx [1]{%
 \ifx #1\expandafter \@firstoftwo
 \else \expandafter \@secondoftwo
 \fi
}%
\providecommand \natexlab [1]{#1}%
\providecommand \enquote  [1]{``#1''}%
\providecommand \bibnamefont  [1]{#1}%
\providecommand \bibfnamefont [1]{#1}%
\providecommand \citenamefont [1]{#1}%
\providecommand \href@noop [0]{\@secondoftwo}%
\providecommand \href [0]{\begingroup \@sanitize@url \@href}%
\providecommand \@href[1]{\@@startlink{#1}\@@href}%
\providecommand \@@href[1]{\endgroup#1\@@endlink}%
\providecommand \@sanitize@url [0]{\catcode `\\12\catcode `\$12\catcode
  `\&12\catcode `\#12\catcode `\^12\catcode `\_12\catcode `\%12\relax}%
\providecommand \@@startlink[1]{}%
\providecommand \@@endlink[0]{}%
\providecommand \url  [0]{\begingroup\@sanitize@url \@url }%
\providecommand \@url [1]{\endgroup\@href {#1}{\urlprefix }}%
\providecommand \urlprefix  [0]{URL }%
\providecommand \Eprint [0]{\href }%
\providecommand \doibase [0]{https://doi.org/}%
\providecommand \selectlanguage [0]{\@gobble}%
\providecommand \bibinfo  [0]{\@secondoftwo}%
\providecommand \bibfield  [0]{\@secondoftwo}%
\providecommand \translation [1]{[#1]}%
\providecommand \BibitemOpen [0]{}%
\providecommand \bibitemStop [0]{}%
\providecommand \bibitemNoStop [0]{.\EOS\space}%
\providecommand \EOS [0]{\spacefactor3000\relax}%
\providecommand \BibitemShut  [1]{\csname bibitem#1\endcsname}%
\let\auto@bib@innerbib\@empty
\bibitem [{\citenamefont {Golding}\ and\ \citenamefont
  {Cox}(2006)}]{gold_2006}%
  \BibitemOpen
  \bibfield  {author} {\bibinfo {author} {\bibfnamefont {I.}~\bibnamefont
  {Golding}}\ and\ \bibinfo {author} {\bibfnamefont {E.~C.}\ \bibnamefont
  {Cox}},\ }\bibfield  {title} {\bibinfo {title} {Physical nature of bacterial
  cytoplasm},\ }\href@noop {} {\bibfield  {journal} {\bibinfo  {journal} {Phys.
  Rev. Lett.}\ }\textbf {\bibinfo {volume} {96}},\ \bibinfo {pages} {098102}
  (\bibinfo {year} {2006})}\BibitemShut {NoStop}%
\bibitem [{\citenamefont {Metzler}\ and\ \citenamefont
  {Klafter}(2000)}]{metzler_2000}%
  \BibitemOpen
  \bibfield  {author} {\bibinfo {author} {\bibfnamefont {R.}~\bibnamefont
  {Metzler}}\ and\ \bibinfo {author} {\bibfnamefont {J.}~\bibnamefont
  {Klafter}},\ }\bibfield  {title} {\bibinfo {title} {The random walk's guide
  to anomalous diffusion: a fractional dynamics approach},\ }\href@noop {}
  {\bibfield  {journal} {\bibinfo  {journal} {Phys. Rep.}\ }\textbf {\bibinfo
  {volume} {339}},\ \bibinfo {pages} {1 } (\bibinfo {year} {2000})}\BibitemShut
  {NoStop}%
\bibitem [{\citenamefont {Weiss}(1994)}]{weiss_1994}%
  \BibitemOpen
  \bibfield  {author} {\bibinfo {author} {\bibfnamefont {G.~H.}\ \bibnamefont
  {Weiss}},\ }\href@noop {} {\emph {\bibinfo {title} {Aspects and
  {A}pplications of the {R}andom {W}alk}}}\ (\bibinfo  {publisher}
  {North-Holland Publishing Co., Amsterdam},\ \bibinfo {year}
  {1994})\BibitemShut {NoStop}%
\bibitem [{\citenamefont {Coffey}\ \emph {et~al.}(2004)\citenamefont {Coffey},
  \citenamefont {Kalmykov},\ and\ \citenamefont {Waldron}}]{Coffey2004}%
  \BibitemOpen
  \bibfield  {author} {\bibinfo {author} {\bibfnamefont {W.~T.}\ \bibnamefont
  {Coffey}}, \bibinfo {author} {\bibfnamefont {Y.~P.}\ \bibnamefont
  {Kalmykov}},\ and\ \bibinfo {author} {\bibfnamefont {J.~T.}\ \bibnamefont
  {Waldron}},\ }\href@noop {} {\emph {\bibinfo {title} {{T}he {L}angevin
  {E}quation}}}\ (\bibinfo  {publisher} {World Scientific Publishing Co. Pte.
  Ltd., Singapore},\ \bibinfo {year} {2004})\BibitemShut {NoStop}%
\bibitem [{\citenamefont {Deng}\ and\ \citenamefont
  {Zhang}(2019)}]{DengZhang2019}%
  \BibitemOpen
  \bibfield  {author} {\bibinfo {author} {\bibfnamefont {W.~H.}\ \bibnamefont
  {Deng}}\ and\ \bibinfo {author} {\bibfnamefont {Z.~J.}\ \bibnamefont
  {Zhang}},\ }\href@noop {} {\emph {\bibinfo {title} {{H}igh {A}ccuracy
  {A}lgorithm for the {D}ifferential {E}quations {G}overning {A}nomalous
  {D}iffusion}}}\ (\bibinfo  {publisher} {World Scientific Publishing Co. Pte.
  Ltd., Singapore},\ \bibinfo {year} {2019})\BibitemShut {NoStop}%
\bibitem [{\citenamefont {Cartea}\ and\ \citenamefont {del
  Castillo-Negrete}(2007)}]{cartea_2007}%
  \BibitemOpen
  \bibfield  {author} {\bibinfo {author} {\bibfnamefont {A.}~\bibnamefont
  {Cartea}}\ and\ \bibinfo {author} {\bibfnamefont {D.}~\bibnamefont {del
  Castillo-Negrete}},\ }\bibfield  {title} {\bibinfo {title} {Fluid limit of
  the continuous-time random walk with general {L}\'evy jump distribution
  functions},\ }\href@noop {} {\bibfield  {journal} {\bibinfo  {journal} {Phys.
  Rev. E}\ }\textbf {\bibinfo {volume} {76}},\ \bibinfo {pages} {041105}
  (\bibinfo {year} {2007})}\BibitemShut {NoStop}%
\bibitem [{\citenamefont {Metzler}\ \emph {et~al.}(2014)\citenamefont
  {Metzler}, \citenamefont {Jeon}, \citenamefont {Cherstvy},\ and\
  \citenamefont {Barkai}}]{metzler_2014}%
  \BibitemOpen
  \bibfield  {author} {\bibinfo {author} {\bibfnamefont {R.}~\bibnamefont
  {Metzler}}, \bibinfo {author} {\bibfnamefont {J.-H.}\ \bibnamefont {Jeon}},
  \bibinfo {author} {\bibfnamefont {A.~G.}\ \bibnamefont {Cherstvy}},\ and\
  \bibinfo {author} {\bibfnamefont {E.}~\bibnamefont {Barkai}},\ }\bibfield
  {title} {\bibinfo {title} {Anomalous diffusion models and their properties:
  non-stationarity{,} non-ergodicity{,} and ageing at the centenary of single
  particle tracking},\ }\href@noop {} {\bibfield  {journal} {\bibinfo
  {journal} {Phys. Chem. Chem. Phys.}\ }\textbf {\bibinfo {volume} {16}},\
  \bibinfo {pages} {24128} (\bibinfo {year} {2014})}\BibitemShut {NoStop}%
\bibitem [{\citenamefont {Fogedby}(1994)}]{Fogedby}%
  \BibitemOpen
  \bibfield  {author} {\bibinfo {author} {\bibfnamefont {H.~C.}\ \bibnamefont
  {Fogedby}},\ }\bibfield  {title} {\bibinfo {title} {Langevin equations for
  continuous time {L}\'evy flights},\ }\href@noop {} {\bibfield  {journal}
  {\bibinfo  {journal} {Phys. Rev. E}\ }\textbf {\bibinfo {volume} {50}},\
  \bibinfo {pages} {1657} (\bibinfo {year} {1994})}\BibitemShut {NoStop}%
\bibitem [{\citenamefont {Magdziarz}\ \emph {et~al.}(2007)\citenamefont
  {Magdziarz}, \citenamefont {Weron},\ and\ \citenamefont
  {Weron}}]{magdziarz_2007}%
  \BibitemOpen
  \bibfield  {author} {\bibinfo {author} {\bibfnamefont {M.}~\bibnamefont
  {Magdziarz}}, \bibinfo {author} {\bibfnamefont {A.}~\bibnamefont {Weron}},\
  and\ \bibinfo {author} {\bibfnamefont {K.}~\bibnamefont {Weron}},\ }\bibfield
   {title} {\bibinfo {title} {Fractional {F}okker-{P}lanck dynamics: Stochastic
  representation and computer simulation},\ }\href@noop {} {\bibfield
  {journal} {\bibinfo  {journal} {Phys. Rev. E}\ }\textbf {\bibinfo {volume}
  {75}},\ \bibinfo {pages} {016708} (\bibinfo {year} {2007})}\BibitemShut
  {NoStop}%
\bibitem [{\citenamefont {Zaburdaev}\ \emph {et~al.}(2015)\citenamefont
  {Zaburdaev}, \citenamefont {Denisov},\ and\ \citenamefont
  {Klafter}}]{zaburdaev:2015}%
  \BibitemOpen
  \bibfield  {author} {\bibinfo {author} {\bibfnamefont {V.}~\bibnamefont
  {Zaburdaev}}, \bibinfo {author} {\bibfnamefont {S.}~\bibnamefont {Denisov}},\
  and\ \bibinfo {author} {\bibfnamefont {J.}~\bibnamefont {Klafter}},\
  }\bibfield  {title} {\bibinfo {title} {L\'{e}vy walks},\ }\href@noop {}
  {\bibfield  {journal} {\bibinfo  {journal} {Rev. Mod. Phys.}\ }\textbf
  {\bibinfo {volume} {87}},\ \bibinfo {pages} {483} (\bibinfo {year}
  {2015})}\BibitemShut {NoStop}%
\bibitem [{\citenamefont {Zaburdaev}\ \emph {et~al.}(2016)\citenamefont
  {Zaburdaev}, \citenamefont {Fouxon}, \citenamefont {Denisov},\ and\
  \citenamefont {Barkai}}]{zaburdaev_2016}%
  \BibitemOpen
  \bibfield  {author} {\bibinfo {author} {\bibfnamefont {V.}~\bibnamefont
  {Zaburdaev}}, \bibinfo {author} {\bibfnamefont {I.}~\bibnamefont {Fouxon}},
  \bibinfo {author} {\bibfnamefont {S.}~\bibnamefont {Denisov}},\ and\ \bibinfo
  {author} {\bibfnamefont {E.}~\bibnamefont {Barkai}},\ }\bibfield  {title}
  {\bibinfo {title} {Superdiffusive dispersals impart the geometry of
  underlying random walks},\ }\href@noop {} {\bibfield  {journal} {\bibinfo
  {journal} {Phys. Rev. Lett.}\ }\textbf {\bibinfo {volume} {117}},\ \bibinfo
  {pages} {270601} (\bibinfo {year} {2016})}\BibitemShut {NoStop}%
\bibitem [{\citenamefont {Zaburdaev}\ \emph {et~al.}(2008)\citenamefont
  {Zaburdaev}, \citenamefont {Schmiedeberg},\ and\ \citenamefont
  {Stark}}]{zaburdaev_2008}%
  \BibitemOpen
  \bibfield  {author} {\bibinfo {author} {\bibfnamefont {V.}~\bibnamefont
  {Zaburdaev}}, \bibinfo {author} {\bibfnamefont {M.}~\bibnamefont
  {Schmiedeberg}},\ and\ \bibinfo {author} {\bibfnamefont {H.}~\bibnamefont
  {Stark}},\ }\bibfield  {title} {\bibinfo {title} {Random walks with random
  velocities},\ }\href@noop {} {\bibfield  {journal} {\bibinfo  {journal}
  {Phys. Rev. E}\ }\textbf {\bibinfo {volume} {78}},\ \bibinfo {pages} {011119}
  (\bibinfo {year} {2008})}\BibitemShut {NoStop}%
\bibitem [{\citenamefont {Xu}\ and\ \citenamefont {Deng}(2018)}]{xu_2018}%
  \BibitemOpen
  \bibfield  {author} {\bibinfo {author} {\bibfnamefont {P.~B.}\ \bibnamefont
  {Xu}}\ and\ \bibinfo {author} {\bibfnamefont {W.~H.}\ \bibnamefont {Deng}},\
  }\bibfield  {title} {\bibinfo {title} {L{\'e}vy walk with multiple internal
  states},\ }\href@noop {} {\bibfield  {journal} {\bibinfo  {journal} {J. Stat.
  Phys.}\ }\textbf {\bibinfo {volume} {173}},\ \bibinfo {pages} {1598}
  (\bibinfo {year} {2018})}\BibitemShut {NoStop}%
\bibitem [{\citenamefont {Dentz}\ \emph {et~al.}(2015)\citenamefont {Dentz},
  \citenamefont {Le~Borgne}, \citenamefont {Lester},\ and\ \citenamefont
  {de~Barros}}]{dentz}%
  \BibitemOpen
  \bibfield  {author} {\bibinfo {author} {\bibfnamefont {M.}~\bibnamefont
  {Dentz}}, \bibinfo {author} {\bibfnamefont {T.}~\bibnamefont {Le~Borgne}},
  \bibinfo {author} {\bibfnamefont {D.~R.}\ \bibnamefont {Lester}},\ and\
  \bibinfo {author} {\bibfnamefont {F.~P.~J.}\ \bibnamefont {de~Barros}},\
  }\bibfield  {title} {\bibinfo {title} {Scaling forms of particle densities
  for {L}\'evy walks and strong anomalous diffusion},\ }\href@noop {}
  {\bibfield  {journal} {\bibinfo  {journal} {Phys. Rev. E}\ }\textbf {\bibinfo
  {volume} {92}},\ \bibinfo {pages} {032128} (\bibinfo {year}
  {2015})}\BibitemShut {NoStop}%
\bibitem [{\citenamefont {Gajda}\ and\ \citenamefont
  {Magdziarz}(2010)}]{janusz:2010}%
  \BibitemOpen
  \bibfield  {author} {\bibinfo {author} {\bibfnamefont {J.}~\bibnamefont
  {Gajda}}\ and\ \bibinfo {author} {\bibfnamefont {M.}~\bibnamefont
  {Magdziarz}},\ }\bibfield  {title} {\bibinfo {title} {Fractional
  {F}okker-{P}lanck equation with tempered $\ensuremath{\alpha}$-stable waiting
  times: Langevin picture and computer simulation},\ }\href@noop {} {\bibfield
  {journal} {\bibinfo  {journal} {Phys. Rev. E}\ }\textbf {\bibinfo {volume}
  {82}},\ \bibinfo {pages} {011117} (\bibinfo {year} {2010})}\BibitemShut
  {NoStop}%
\bibitem [{\citenamefont {Sandev}\ \emph {et~al.}(2018)\citenamefont {Sandev},
  \citenamefont {Deng},\ and\ \citenamefont {Xu}}]{trifce:2018}%
  \BibitemOpen
  \bibfield  {author} {\bibinfo {author} {\bibfnamefont {T.}~\bibnamefont
  {Sandev}}, \bibinfo {author} {\bibfnamefont {W.~H.}\ \bibnamefont {Deng}},\
  and\ \bibinfo {author} {\bibfnamefont {P.~B.}\ \bibnamefont {Xu}},\
  }\bibfield  {title} {\bibinfo {title} {Models for characterizing the
  transition among anomalous diffusions with different diffusion exponents},\
  }\href@noop {} {\bibfield  {journal} {\bibinfo  {journal} {J. Phys. A: Math.
  Theor.}\ }\textbf {\bibinfo {volume} {51}},\ \bibinfo {pages} {405002}
  (\bibinfo {year} {2018})}\BibitemShut {NoStop}%
\bibitem [{\citenamefont {Kr\"usemann}\ \emph {et~al.}(2014)\citenamefont
  {Kr\"usemann}, \citenamefont {Godec},\ and\ \citenamefont
  {Metzler}}]{krusemann}%
  \BibitemOpen
  \bibfield  {author} {\bibinfo {author} {\bibfnamefont {H.}~\bibnamefont
  {Kr\"usemann}}, \bibinfo {author} {\bibfnamefont {A.~c.~v.}\ \bibnamefont
  {Godec}},\ and\ \bibinfo {author} {\bibfnamefont {R.}~\bibnamefont
  {Metzler}},\ }\bibfield  {title} {\bibinfo {title} {First-passage statistics
  for aging diffusion in systems with annealed and quenched disorder},\
  }\href@noop {} {\bibfield  {journal} {\bibinfo  {journal} {Phys. Rev. E}\
  }\textbf {\bibinfo {volume} {89}},\ \bibinfo {pages} {040101} (\bibinfo
  {year} {2014})}\BibitemShut {NoStop}%
\bibitem [{\citenamefont {Deng}\ \emph {et~al.}(2017)\citenamefont {Deng},
  \citenamefont {Wu},\ and\ \citenamefont {Wang}}]{deng:2017}%
  \BibitemOpen
  \bibfield  {author} {\bibinfo {author} {\bibfnamefont {W.~H.}\ \bibnamefont
  {Deng}}, \bibinfo {author} {\bibfnamefont {X.~C.}\ \bibnamefont {Wu}},\ and\
  \bibinfo {author} {\bibfnamefont {W.~L.}\ \bibnamefont {Wang}},\ }\bibfield
  {title} {\bibinfo {title} {Mean exit time and escape probability for the
  anomalous processes with the tempered power-law waiting times},\ }\href@noop
  {} {\bibfield  {journal} {\bibinfo  {journal} {EPL}\ }\textbf {\bibinfo
  {volume} {117}},\ \bibinfo {pages} {10009} (\bibinfo {year}
  {2017})}\BibitemShut {NoStop}%
\bibitem [{\citenamefont {Dybiec}\ and\ \citenamefont
  {Sokolov}(2015)}]{dybiec}%
  \BibitemOpen
  \bibfield  {author} {\bibinfo {author} {\bibfnamefont {B.}~\bibnamefont
  {Dybiec}}\ and\ \bibinfo {author} {\bibfnamefont {I.~M.}\ \bibnamefont
  {Sokolov}},\ }\bibfield  {title} {\bibinfo {title} {Estimation of the
  smallest eigenvalue in fractional escape problems: Semi-analytics and fits},\
  }\href@noop {} {\bibfield  {journal} {\bibinfo  {journal} {Comput. Phys.
  Comm.}\ }\textbf {\bibinfo {volume} {187}},\ \bibinfo {pages} {29} (\bibinfo
  {year} {2015})}\BibitemShut {NoStop}%
\bibitem [{\citenamefont {A.~P.~Prudnikov}\ and\ \citenamefont
  {Marichev}(1990)}]{prudnikov}%
  \BibitemOpen
  \bibfield  {author} {\bibinfo {author} {\bibfnamefont {Y.~A.~B.}\
  \bibnamefont {A.~P.~Prudnikov}}\ and\ \bibinfo {author} {\bibfnamefont
  {O.~I.}\ \bibnamefont {Marichev}},\ }\href@noop {} {\emph {\bibinfo {title}
  {{I}ntegral and {S}eries}}}\ (\bibinfo  {publisher} {Gordon and Breach
  Science Publishers, New York},\ \bibinfo {year} {1990})\BibitemShut {NoStop}%
\bibitem [{\citenamefont {Abramowitz}\ and\ \citenamefont
  {Stegun}(1972)}]{abra}%
  \BibitemOpen
  \bibfield  {author} {\bibinfo {author} {\bibfnamefont {M.}~\bibnamefont
  {Abramowitz}}\ and\ \bibinfo {author} {\bibfnamefont {I.~A.}\ \bibnamefont
  {Stegun}},\ }\href@noop {} {\emph {\bibinfo {title} {{H}andbook of
  {M}athematical {F}unctions with {F}ormulas, {G}raphs, and {M}athematical
  {T}ables}}}\ (\bibinfo  {publisher} {United States Department of Commerce,
  Washington D.C.},\ \bibinfo {year} {1972})\BibitemShut {NoStop}%
\end{thebibliography}%

%
%

\end{document}